\documentclass[a4paper,11pt]{article}
\pdfoutput=1 

\usepackage{jheppub} 
\usepackage{bbm,bm,graphicx,mathtools,color,slashed,hyperref}
\usepackage{amssymb,amsmath}
\usepackage{braket}
\usepackage{subfig}

\newcommand{\diff}{\mathrm{d}}

\newcommand{\tr}{\mathrm{tr}}
\newcommand{\im}{\mathrm{i}}

\newcommand{\rme}{\mathrm{e}}

\newcommand{\ee}{\,\mathrm{e}}

\newcommand{\ZZ}{\mathbb{Z}}
\newcommand{\RR}{\mathbb{R}}

\definecolor{darkred}{rgb}{0.7, 0.0, 0.0}
\newcommand\hl[1]{{\color{black} #1}}
 
\preprint{YITP-24-136}

\title{
Monopole-vortex continuity of $ {\mathcal N}=1 $ super Yang-Mills theory on $\mathbb{R}^2 \times S^1 \times S^1$ with 't~Hooft twist
}

\author[1]{Yui Hayashi,}
\emailAdd{yui.hayashi@yukawa.kyoto-u.ac.jp}
\affiliation[1]{Yukawa Institute for Theoretical Physics, Kyoto University, Kyoto 606-8502, Japan}

\author[2,3]{Tatsuhiro Misumi,}
\emailAdd{misumi@phys.kindai.ac.jp}
\affiliation[2]{Department of Physics, Kindai University, Osaka 577-8502, Japan}
\affiliation[3]{Research and Education Center for Natural Sciences, Keio University, Kanagawa 223-8521, Japan}

\author[1]{Yuya Tanizaki}
\emailAdd{yuya.tanizaki@yukawa.kyoto-u.ac.jp}

\abstract{
We study ${\mathcal N} = 1$ $SU(N)$ super Yang-Mills (SYM) theory on $\mathbb{R}^2\times (S^1)_3\times (S^1)_4$ with the 't Hooft twist. 
The theory becomes weakly coupled if the length $L_4$ of $(S^1)_4$ is sufficiently small, $NL_4\Lambda\ll 1$. We explore the nonperturbative dynamics at the weak-coupling regime by changing the size of $L_3$ and uncover how $3$d monopole/bion-based effective theory for $L_3\gg L_4$ is related to the $2$d vortex-based theory for $L_3\approx L_4$. 
The highlights of our results are 
(1)~the smooth ``weak-weak" continuity of the vacuum structure and gluino condensate during the $3$d-$2$d dimensional reduction, 
(2)~the \hl{switching} of Wilson loop behavior from the area law in $3$d to the perimeter law in $2$d via a ``double-string" picture, 
(3)~the role of mass deformation in breaking discrete chiral symmetry and restoring the area law in $2$d, and 
(4)~the microscopic investigation of bions during the reduction from $3$d to $2$d and the cancellation of the vacuum energy due to the hidden topological angle. 
We also discuss the generalization of our results for (1)--(3) from $\mathcal{N}=1$ SYM to QCD with adjoint quarks.  
}

\begin{document}

\maketitle


\section{Introduction}

Understanding nonperturbative properties of $4$d Quantum Chromodynamics (QCD) and Yang-Mills theory remains one of the most challenging and fundamental problems of particle physics. To study these properties, it is often useful to consider various deformations of strongly-coupled field theories to weakly-coupled ones while keeping their qualitative features. After taking suitable deformations, we can apply weak-coupling analyses to extract some properties of the confinement physics.

Because of the asymptotic freedom, 4d gauge theories become weakly coupled when put on small compactified spacetime. Although the deconfinement transition happens if this is done naively, it has been uncovered that confinement is kept intact under suitable setups. Currently, there are two ways to achieve it, and monopoles and center vortices play essential roles in those setups to explain confinement. 
One way is to consider Yang-Mills theory with the double-trace deformation (or QCD with adjoint quarks) on $ \mathbb{R}^3 \times S^1 $~\cite{Davies:1999uw, Davies:2000nw, Unsal:2007vu, Unsal:2007jx, Shifman:2008ja} (see \cite{Poppitz:2021cxe} for a review). 
For sufficiently small $S^1$, the theory becomes abelianized, and the monopole (or monopole-molecule) plays pivotal roles in confinement and other nonperturbative phenomena. 
The other way is to consider Yang-Mills theory or QCD on $ \mathbb{R}^2 \times T^2 $ with 't Hooft flux. For small $T^2$, we can derive two-dimensional theories of center vortices to explain confinement and also chiral effective Lagrangian~\cite{Tanizaki:2022ngt, Tanizaki:2022plm, Hayashi:2023wwi, Hayashi:2024qkm, Hayashi:2024gxv}. Further compactification of this setup to $ \mathbb{R}^1 \times S^1 \times T^2 $ enables the analysis within the framework of quantum mechanics~\cite{Yamazaki:2017ulc, Cox:2021vsa}. 
Semiclassical analyses of these compactified setups facilitate the study of confinement and chiral symmetry breaking in a controlled setting and uncover qualitative behaviors of confinement vacua. These compactifications should achieve the adiabatic continuity between weakly-coupled low-dimensional theories and strongly-coupled original theories, while its full demonstration is still an important task.

We now have two distinct semiclassical descriptions on $\mathbb{R}^3\times S^1$ and $\mathbb{R}^2\times T^2$ with 't~Hooft flux. In both setups, the fractional instantons play important roles in explaining confinement, but their realizations are different between these two cases: they take the form of magnetic monopoles on $\mathbb{R}^3\times S^1$ while they become center vortices on $\mathbb{R}^2\times T^2$.   
In recent work by some of the present authors, we investigated the connection between magnetic monopoles and center vortices and unified these two semiclassical pictures~\cite{Hayashi:2024yjc} (see also \cite{Guvendik:2024umd}).
Using the dual photon effective theory for the 3d semiclassical analysis, we showed how the 2d center-vortex theory naturally appears via the center-symmetry twisted boundary condition and identified monopoles as the junction of the center-vortex configuration. 
This unification bridges the gap between monopole-based and vortex-based semiclassical frameworks, which leads to the weak-weak continuity of the confinement mechanism.

In this paper, we extend the previous study to understand the behavior of ${\mathcal N}=1$ $SU(N)$ super Yang-Mills (SYM) theory under compactification from $ {\mathbb R}^3 \times S^1 $ to $\mathbb{R}^2 \times T^2 $ with the 't Hooft twist. 
The $\mathcal{N}=1$ SYM theory serves as the ideal prototype realizing the adiabatic continuity of the confinement vacua: 
Certain observables, such as the Witten index~\cite{Witten:1982df, Witten:2000nv}, enjoy the topological nature thanks to supersymmetry, and we can apply the semiclassical weak-coupling analysis for those quantities. 
In the ${\mathcal N}=1$ $SU(N)$ SYM theory, significant progress has been made in the monopole semiclassics on $\mathbb{R}^3 \times S^1$, but a clear understanding of the center-vortex semiclassics on $\mathbb{R}^2 \times T^2 $  is still lacking at the quantitative level as will be discussed at the end of Section~\ref{sec:review_Bion_Vortex}.
Therefore, in this paper, we study the weak-weak continuity, or the monopole-vortex continuity, in the SYM theory to understand the confinement mechanism across different dimensional compactifications. 
The following is the highlight of this paper.

{\it Monopole and center-vortex semiclassics}: 
In Section~\ref{sec:review_Bion_Vortex}, we begin with reviewing the semiclassical approaches on $ {\mathbb R}^3 \times S^1 $ and $ {\mathbb R}^2 \times T^2 $, following Refs.~\cite{Davies:1999uw, Davies:2000nw, Unsal:2007vu,Unsal:2007jx} and Ref.~\cite{Tanizaki:2022ngt}, respectively. 
On $\mathbb{R}^3 \times S^1 $, the 3d gauge field is abelianized as 
$SU(N) \xrightarrow{\mathrm{Higgs}} U(1)^{N-1}$ at generic points of classical moduli, and the fundamental monopoles carry two fermionic zero modes as they have $1/N$ fractional topological charge. 
The nonperturbative contribution to the bosonic potential comes from monopole molecules called ``magnetic bions" and ``neutral bions", which yield the mass gap to the dual photon and holonomy fields, respectively. 
On $ {\mathbb R}^2 \times T^2 $ with 't Hooft flux, the 2d gauge field has a perturbative gap due to the boundary condition, and ``center vortex" describes the tunneling event that carries two fermionic zero modes. 
Therefore, the center vortex generates the gluino condensate but does not produce the area law of Wilson loops, which is consistent with the mixed 't Hooft anomaly after the 2d reduction. 

{\it ``Weak-weak" continuity of the vacuum structure}:
In Section~\ref{sec:WeakWeakContinuity}, we consider ${\mathcal N}=1$ $SU(N)$ SYM on $ {\mathbb R}^2 \times (S^{1})_3 \times (S^{1})_4$ with the 't~Hooft twist and vary the size $L_3$ while keeping $N\Lambda L_4\ll 1$. We elucidate that the twisted boundary condition originating from the 't Hooft twist guarantees the transition from monopoles to center-vortices, and we establish the ``weak-weak'' continuity of the vacuum structure, such as the superpotential and gluino condensate. We also study the implication of the mass deformation on the vacuum structure and the generalization of these arguments to QCD(adj.).

{\it Double string picture and the \hl{switching} from 3d area law to 2d perimeter law}:
In Section~\ref{sec:DoubleStringPicture}, we investigate the behavior of the Wilson loop during the dimensional transition from $ {\mathbb R}^3 \times S^1 $ to $ {\mathbb R}^2 \times T^2 $ with 't~Hooft flux. 
The 4d mixed anomaly between the center and chiral symmetry produces the 2d mixed anomaly between the 1-form center and chiral symmetry under the presence of 't~Hooft flux, which requires the 2d perimeter law of the Wilson loop. 
We show that the perimeter law of the 2d center-vortex semiclassics can be understood from the 3d monopole through the ``double-string picture".
With the double-string picture, the Wilson loop creates the domain wall for the broken chiral symmetry that wraps along the $S^1$ direction, and the Wilson loop serves as the junction of the domain wall. 
With the mass deformation, Wilson loops obey the area law due to the absence of the discrete chiral symmetry.

{\it Bion contribution and vacuum energy}:
In Section~\ref{sec:BionContribution}, we examine the fate of the bion-induced confinement during the dimensional reduction from 3d to 2d.  
The calculation of the bion amplitude shows that magnetic bions disappear under the compactification. Although the magnetic bions exist as the local minima for $L_3 \gg L_4 /g^2$, those local minima merge with other saddle points around $L_3 \sim L_4 /g^2$ so that only the unstable saddle remains as their remnant. 
We interpret this observation as a microscopic signature of the crossover from the 3d monopole/bion picture to the 2d center-vortex picture.
We also discuss how the vacuum energy continues to vanish during the reduction from 3d to 2d in terms of the bion amplitude,
and we find that its cancellation occurs in a parallel manner to that in the $2$d $\mathcal{N}=(2,2)$ supersymmetric ${\mathbb C}P^{1}$ model on $\mathbb{R}\times S^1$ with the flavor-twisted boundary condition~\cite{Fujimori:2016ljw}.


\section{Review on \texorpdfstring{$ \mathbb{R}^3 \times S^1 $}{R3xS1} semiclassics and \texorpdfstring{$ \mathbb{R}^2 \times T^2 $}{R2xT2} semiclassics}
\label{sec:review_Bion_Vortex}

In this section, we review the two semiclassical approaches for $\mathcal{N}=1$ $SU(N)$ super Yang-Mills (SYM) theory: $\mathbb{R}^3 \times S^1$ semiclassics~\cite{Davies:1999uw, Davies:2000nw, Unsal:2007vu, Unsal:2007jx} and $ \mathbb{R}^2 \times T^2 $ semiclassics~\cite{Tanizaki:2022ngt}.

The $\mathcal{N}=1$ $SU(N)$ SYM theory is the $4$d $SU(N)$ Yang-Mills theory with one massless adjoint Weyl fermion $\lambda$, called the gluino.
The action is,
\begin{align}
    S= \frac{1}{g^2} \int \operatorname{tr}(f \wedge \star f) - \frac{\im \theta}{8 \pi^2} \int \operatorname{tr}(f \wedge f) + \frac{2 \im }{g^2} \int \diff^4x ~\operatorname{tr}(\Bar{\lambda} \Bar{\sigma}^\mu D_\mu \lambda),
\end{align}
where $a=a_\mu \diff x^\mu$ is the $SU(N)$ gauge field, $f = \diff a + \im\, a \wedge a$ is its field strength, and $D_\mu \lambda = \partial_\mu \lambda + \im\, [a_\mu, \lambda]$ is the covariant derivative in the adjoint representation, and this theory enjoys the $4$d $\mathcal{N}=1$ supersymmetry. 
This theory has the $\mathbb{Z}_N$ $1$-form symmetry, denoted as $\mathbb{Z}_N^{[1]}$ and called the center symmetry, and also has the $\mathbb{Z}_{2N}$ discrete chiral symmetry, $\lambda \mapsto \rme^{2\pi\im/(2N)}\lambda$. 
There is the mixed 't~Hooft anomaly between $\mathbb{Z}_N^{[1]}$ and $(\mathbb{Z}_{2N})_{\mathrm{chiral}}$, and the anomaly matching claims the spontaneous breaking of chiral symmetry, $(\mathbb{Z}_{2N})_{\mathrm{chiral}}\xrightarrow{\mathrm{SSB}} \mathbb{Z}_2$, if the system is in the confinement phase. 
We note that this $\mathbb{Z}_2$ subgroup is the fermion parity, and it cannot be spontaneously broken for Lorentz-invariant vacua. 

The partition function on $\mathbb{R}^3\times S^1$ with the periodic boundary condition for $\lambda$ is called the Witten index, which is topologically protected and gives $N$ at any size of $S^1$~\cite{Witten:1982df}. 
This $N$-fold degeneracy can be naturally understood as the result of the discrete chiral symmetry breaking associated with the chiral condensate $\langle \tr(\lambda \lambda)\rangle$, and it satisfies the anomaly matching constraint. 
Due to its topological nature, we can deform the theory into a weakly coupled one without any phase transition through the compactification.

\subsection{3d semiclassics: \texorpdfstring{$ \mathbb{R}^3 \times S^1 $}{R3xS1} with spatial compactification}
\label{sec:MonopoleSemiclassics}

Let us first consider the $\mathcal{N}=1$ SYM on $\mathbb{R}^3\times S^1$, where the size $L_4$ of $S^1$ is sufficiently small compared with the strong scale $1/\Lambda$, 
\begin{equation}
    N \Lambda L_4\ll 1. 
\end{equation}
We take the periodic boundary condition for $\lambda$ (the spatial compactification), and this preserves the supersymmetry. 
Here, the 3d effective theory at small $S^1$ is briefly reviewed based on \cite{Davies:1999uw, Davies:2000nw, Unsal:2007vu, Unsal:2007jx}.

At small $S^1$, the bosonic degrees of freedom in the effective theory are the 3d gauge field and holonomy $P_4=\mathcal{P}\exp(\im \int_{S^1} a_4 \diff x_4)$ along $S^1$ (Polyakov loop). 
For pure YM or SYM with the antiperiodic fermion boundary condition, the one-loop Gross-Pisarski-Yaffe (GPY) potential for the holonomy is generated~\cite{Gross:1980br}, and the holonomy gets the vacuum expectation value that violates the center symmetry. 
For $\mathcal{N} = 1$ SYM theory with the periodic fermion boundary condition, the perturbative GPY potential exactly cancels between the gluon and gluino contributions due to supersymmetry, and the moduli space for $P_4$ remains to be flat within the perturbative computation.
Thus, both 3d gauge field and holonomy survive as the low-energy degrees of freedom.

Let us take the Polyakov gauge that diagonalizes the holonomy, then $P_4$ becomes
\begin{align}
    P_4 = \operatorname{diag}(\rme^{\im \varphi_1}, \cdots, \rme^{\im \varphi_N} )\,,
    \label{eq:PolyakovGauge}
\end{align}
with the condition $\varphi_1 + \cdots + \varphi_N = 0 ~(\operatorname{mod} 2\pi)$. 
It is convenient to denote them as the $N$-component field, $\Vec{\phi}=(\varphi_1,\ldots, \varphi_{N-1},-\varphi_1-\cdots-\varphi_{N-1})$, whose periodicity is 
\begin{align}
    \Vec{\phi} \sim \Vec{\phi} + 2 \pi \vec{\alpha}_i,
\end{align}
where $\vec{\alpha}_i~(i = 1,\cdots,N-1)$ is the positive simple root.\footnote{Our convention for the $SU(N)$ root and weight vectors is summarized in Appendix~\ref{appendix:RootWeightConvention}. We can expand $\Vec{\phi}$ as $\Vec{\phi}=\sum_i \phi_i \Vec{\alpha}_i$ by taking $\varphi_1=\phi_1$, $\varphi_2=\phi_2-\phi_1$, \ldots, $\varphi_{N-1}=\phi_{N-1}-\phi_{N-2}$, and $\varphi_N=-\phi_{N-1}$. }

With the holonomy background \eqref{eq:PolyakovGauge}, the 3d gluon field $(a^{ij})_{i,j=1,\ldots, N}$ gets the Kaluza-Klein mass $\frac{\varphi_i-\varphi_j}{L_4}$ mod $\frac{2\pi}{L_4}\mathbb{Z}$, and the diagonal gluons are gapless while off-diagonal ones acquire the nonzero gap at generic values of $P_4$.\footnote{The same is true also for gluinos, and we denote the gapless diagonal components as $\Vec{\lambda}$. } 
Therefore, the 3d gauge field is abelianized at almost all holonomies, $SU(N) \xrightarrow{\mathrm{Higgs}} U(1)^{N-1}$. By using the 3d Abelian duality, we can rewrite this 3d $U(1)^{N-1}$ gauge field into a $U(1)^{N-1}$-valued scalar field, called the dual photon $\vec{\sigma}$.
Because of the electromagnetic duality, its periodicity is 
\begin{align}
    \vec{\sigma} \sim \vec{\sigma} + 2 \pi \vec{\mu}_k,
\end{align}
where $\vec{\mu}_k ~(k=1,\cdots, N-1)$ denotes the fundamental weight.
Noting that the permutation redundancy remains, the target space of compact bosons $(\vec{\phi},\vec{\sigma})$ is given by
\begin{align}
    (\vec{\phi},\vec{\sigma}) \in \frac{\mathbb{R}^{N-1}/2\pi \Lambda_{\mathrm{roots}} \times \mathbb{R}^{N-1}/2\pi \Lambda_{\mathrm{weights}}}{S_N},
\end{align}
where $S_N$ denotes the Weyl permutation group, and $\Lambda_{\mathrm{roots}}$ and $\Lambda_{\mathrm{weights}}$ are the root and weight lattices, respectively.

One can eliminate the permutation redundancy by restricting the space of $\vec{\phi}$ to the Weyl chamber, e.g.,
\begin{align}
    \vec{\alpha}_i \cdot \vec{\phi} > 0,~~ -\vec{\alpha}_N \cdot \vec{\phi} <2 \pi,
\end{align}
where $\vec{\alpha}_N = - (\vec{\alpha}_1 + \cdots + \vec{\alpha}_{N-1})$ is the Affine simple root. 
In this gauge, the center transformation is represented by~\cite{Anber:2015wha} 
\begin{align}
    (\vec{\phi} , \vec{\sigma} ) \mapsto (P_{\mathrm{W}} \vec{\phi} + 2 \pi \vec{\mu}_1, P_{\mathrm{W}} \vec{\sigma} ), \label{eq:3d-center-transf}
\end{align}
where $P_{\mathrm{W}}$ denotes the cyclic Weyl permutation.
The center-symmetric holonomy $\vec{\phi}_c$ can be written as,
\begin{align}
    \vec{\phi}_c = \frac{2 \pi}{N} \vec{\rho} = \frac{2 \pi}{N} (\vec{\mu}_1 + \cdots + \vec{\mu}_{N-1}),
    \label{eq:CenterSymmetricHolonomy}
\end{align}
where $\vec{\rho}$ is called the Weyl vector. This center-symmetric holonomy corresponds to $P_4=C$, up to gauge, where the clock matrix is given by $C = \rme^{\frac{\im (N+1) \pi}{N}} \operatorname{diag} (1, \rme^{\frac{2 \pi \im}{N}}, \cdots, \rme^{\frac{2 \pi \im (N-1)}{N}} )$. 

Here, we introduce the complex scalar variable:
\begin{align}
    \vec{z} := \im  \left[ \vec{\sigma} + \left( \frac{\theta}{2 \pi} + \frac{4 \pi \im}{g^2} \right) (\vec{\phi} - \vec{\phi}_c) \right]. \label{eq:complex_var_Z}
\end{align}
With this notation, the tree-level kinematic term of the bosonic degrees of freedom can be neatly written as,
\begin{align}
    S_{\mathrm{3d, bosonic}} = \frac{1}{g^2 L_4} |\diff \vec{\phi}|^2 + \frac{g^2}{16 \pi^2 L_4} \left| \diff \vec{\sigma} + \frac{\theta}{2 \pi } \diff \vec{\phi}  \right|^2 = \frac{g^2}{16 \pi^2 L_4} \left| \diff \vec{z}  \right|^2.
\end{align}
We note that the complex scalar $\vec{z}$ and the Cartan gluino $\Vec{\lambda}$ forms the 3d $\mathcal{N}=2$ supersymmetric multiplet, and $\vec{z}$ is the lowest component of the superfield, $\vec{Z}=\vec{z}+\sqrt{2}\vartheta \Vec{\lambda}+\cdots$, in the $4d$ $\mathcal{N}=1$ chiral multiplet notation. Therefore, this neat feature of the effective Lagrangian in terms of $\vec{z}$ continues even after taking into account the loop corrections. 

For confinement and chiral symmetry breaking, we must go beyond perturbation theory and need nonperturbative objects; monopoles and bions.
There are $N$-types of fundamental monopoles in this setup: $N-1$ Bogomol'nyi-Prasad-Sommerfield (BPS) monopoles and one Kaluza-Klein (KK) monopole~\cite{Lee:1997vp, Lee:1998bb, Lee:1998vu, Kraan:1998kp, Kraan:1998pm, Kraan:1998sn}.
Magnetic charges of BPS monopoles are simple roots $\{ \vec{\alpha}_i \}_{i = 1, \cdots , N-1}$, and the magnetic charge of the KK monopole is the affine root $\vec{\alpha}_N$.
All these monopoles carry two fermionic zero modes according to the Callias index theorem, and one can find the monopole-instanton vertex,
\begin{align}
    [\mathcal{M}_i] \sim \rme^{- \frac{8 \pi^2}{Ng^2} + \frac{\im \theta}{N}} \rme^{\vec{\alpha}_i \cdot \vec{z}} (\vec{\alpha}_i \cdot \vec{\lambda})^2 ~~~(i = 1 , \cdots, N). \label{eq:monopole-vertex}
\end{align}
For the anti-monopoles, we assign
\begin{align}
    [\mathcal{M}_i^*] \sim \rme^{- \frac{8 \pi^2}{Ng^2} - \frac{\im \theta}{N}} \rme^{\vec{\alpha}_i \cdot \vec{z}^*} (\vec{\alpha}_i \cdot \vec{\Bar{\lambda}})^2 ~~~(i = 1 , \cdots, N).
\end{align}
The monopole vertex should be invariant under the discrete chiral transformation, and thus it acts not only on Cartan gluinos but also on the complex scalar:
\begin{equation}
    (\mathbb{Z}_{2N})_{\mathrm{chiral}}: \vec{\lambda}\mapsto \rme^{\frac{2\pi \im}{2N}}\vec{\lambda}, \quad \vec{z}\mapsto \vec{z}-\frac{2\pi\im}{N}\vec{\rho}. 
\end{equation}
Therefore, the vacuum expectation value of $\rme^{\vec{\alpha}_i\cdot \vec{z}}$ plays the role of the order parameter for the discrete chiral symmetry breaking. 

Since the monopoles carry fermionic zero modes, the leading contributions for the bosonic nonperturbative potential come from bions, monopole-antimonopole molecules.
The bion amplitude should be roughly given by
\begin{align}
    [\mathcal{M}_i \mathcal{M}^*_j] \sim \rme^{- 2 \frac{8 \pi^2}{Ng^2}} \rme^{\vec{\alpha}_i \cdot \vec{z} + \vec{\alpha}_j \cdot \vec{z}^* } (\vec{\alpha}_i \cdot \vec{\alpha}_j)^2,
\end{align}
Note that only $[\mathcal{M}_i \mathcal{M}^*_j]$ with $j = i-1,i,i+1$ are nonzero (corresponding to the nonvanishing Cartan matrix elements).
The bion $[\mathcal{M}_i \mathcal{M}^*_j]$ with $j = i \pm 1$ is called magnetic bion, and $[\mathcal{M}_i \mathcal{M}^*_i]$ is called neutral bion.

The monopole and bion contributions to the effective Lagrangian are tightly constrained by supersymmetry. 
They can be summarized as the affine Toda superpotential, which is obtained from calculating the monopole amplitude~\cite{Davies:1999uw, Davies:2000nw}, 
\begin{align} 
    W(\vec{Z}) = \frac{ M_{PV}^3 L_4 \rme^{-\frac{8 \pi^2}{Ng^2} + \frac{\im \theta}{N}}}{g^2} \sum_{i=1}^{N} \rme^{\vec{\alpha}_i \cdot \vec{Z}} = L_4 \Lambda^3 \rme^{ \frac{\im \theta}{N}} \sum_{i=1}^{N} \rme^{\vec{\alpha}_i \cdot \vec{Z}},
    \label{eq:AffineTodaW}
\end{align}
where $M_{PV}$ is the Pauli-Villars regulator, and we defined $\Lambda^3 =  \frac{1}{g^2} M_{PV}^3 \rme^{-\frac{8 \pi^2}{Ng^2}}$ 
(matched to the NSVZ scheme after changing to the canonical gauge coupling from the holomorphic one).
This superpotential not only describes the monopole-induced interactions but also automatically leads to the bion-induced bosonic potential:
\begin{align} 
V(\vec{z}) &= \frac{16\pi^2 L_4}{g^2} \left|\frac{\partial W(\vec{z})}{\partial \vec{z}}\right|^2= V_0\sum_{i,j=1}^{N}\rme^{\vec{\alpha}_i \cdot \vec{z} + \vec{\alpha}_j \cdot \vec{z}^* } (\vec{\alpha}_i \cdot \vec{\alpha}_j) \notag\\
&=V_0 \sum_{i=1}^N \left| \rme^{\vec{\alpha}_i \cdot \vec{z}} - \rme^{\vec{\alpha}_{i-1} \cdot \vec{z}}  \right|^2, 
\label{eq:BionPotential}
\end{align}
where $V_0=\frac{16\pi^2}{g^2}\Lambda^6 L_4^3$. 
Usually, the semiclassical objects lower the effective potential, and magnetic bions follow this rule. 
However, neutral bions give the opposite contribution and this relative sign is important to have supersymmetric vacua. 
In terms of the Lefschetz-thimble integration over the quasi-moduli, this relative sign is interpreted as the ``hidden topological angle''~\cite{Behtash:2015kna, Behtash:2015zha}, and we will discuss it in more detail in Section~\ref{sec:BionContribution}.

Minimization of the bosonic potential is achieved by requiring that $\rme^{\vec{\alpha}_i\cdot \vec{z}}=\rme^{\vec{\alpha}_j\cdot \vec{z}}$ for all $i,j=1,\ldots, N$. Since $\prod_{i=1}^{N}\rme^{\vec{\alpha}_i\cdot\vec{z}}=1$, we can easily find the $N$-fold degenerate vacua,
\begin{align} 
&\rme^{\vec{\alpha}_i\cdot \vec{z}}=\rme^{2\pi \im k/N } ~~~ (i=1,\ldots, N)\notag\\
&\Leftrightarrow \quad 
\vec{z} = \frac{2 \pi k \vec{\rho}}{N} \im , 
\label{eq:vacua_in_Z}
\end{align}
for $k=0,1,\ldots, N-1$, which spontaneously break the discrete chiral symmetry. 
The bosonic potential produces the mass gap to the complex scalar $\vec{z}$, and if we look more closely, the dual photon gets a mass due to magnetic bions, while the holonomy field gets a mass due to neutral bions~\cite{Unsal:2007vu, Unsal:2007jx}. This is the semiclassical confinement mechanism from monopoles and bions for the $4$d $\mathcal{N}=1$ SYM on $\mathbb{R}^3\times S^1$.

\subsection{2d semiclassics: \texorpdfstring{$ \mathbb{R}^2 \times T^2 $}{R2xT2} with 't Hooft flux}
\label{sec:VortexSemiclassics}

The other semiclassical method is to compactify $\mathbb{R}^4$ to $\mathbb{R}^2 \times T^2$ with 't Hooft flux, and let us give a brief review based on \cite{Tanizaki:2022ngt} (see also \cite{Yamazaki:2017ulc, Cox:2021vsa}).  
We set the size of two cycles to be the same, $L_3=L_4=L$, and they are supposed to be sufficiently small, $N\Lambda L\ll 1$. 

To form the torus $T^2$, we impose the identification $(x_3,x_4) \sim (x_3+L,x_4) \sim (x_3,x_4+L)$.
Since the $\mathcal{N}=1$ SYM contains only adjoint fields, the 't Hooft flux can be introduced.
We turn on the unit 't Hooft flux by choosing the $SU(N)$-valued transition functions $g_3(x_4)$ and $g_4(x_3)$ such that
\begin{align}
    g_3(L)^\dagger g_4(0)^\dagger g_3(0) g_4(L) = \rme^{\frac{2 \pi \im}{N}} 1_{N \times N}.
\end{align}
The 't Hooft flux corresponds to the background gauge field for the $\mathbb{Z}_N^{[1]}$ symmetry along the compactified $3$-$4$ directions. 
We can perform the gauge transformation so that these transition functions become the clock matrix $C$ and shift matrix $S$ of $SU(N)$:
\begin{align}
    g_3(x_4) = S,~~~~g_4(x_3) = C, \label{eq:transition_functions_gauge}
\end{align}
where $C =\rme^{\im \alpha} \operatorname{diag}(1,\rme^{\frac{2 \pi \im}{N}}, \cdots,\rme^{\frac{2 \pi \im (N-1)}{N}})$, $(S)_{ij} =\rme^{\im \alpha} \delta_{i+1,j}$, and $\rme^{\im N \alpha}=(-1)^{N+1}$.
This gauge choice turns out to be useful for the analysis on small $T^2$.

Under the $T^2$ compactification, the 1-form symmetry $\mathbb{Z}_N^{[1]}$ in the 4d spacetime becomes, in terms of the 2d description, 
\begin{align}
    \left( \mathbb{Z}_N^{[1]} \right)_{\mathrm{4d}} \rightarrow \left( \mathbb{Z}_N^{[1]} \times  \left(\mathbb{Z}_N^{[0]} \times \mathbb{Z}_N^{[0]} \right) \right)_{\mathrm{2d}}, 
\end{align}
where $\left( \mathbb{Z}_N^{[0]} \times \mathbb{Z}_N^{[0]} \right)_{\mathrm{2d}}$ is the center symmetry acting on the Polyakov loops $P_3, P_4$ along each cycle of $T^2$, and $\left(  \mathbb{Z}_N^{[1]} \right)_{\mathrm{2d}}$ is the 1-form center symmetry acting on the spatial Wilson loop on $\mathbb{R}^2$.
At small $T^2$, the gauge field becomes flat along the $3$-$4$ direction, $f_{34}=0$. 
In the gauge (\ref{eq:transition_functions_gauge}), the boundary conditions for the gluon and gluino are given by
\begin{align}
    &a_{\mu} (\Vec{x},x_3+L,x_4) = S a_{\mu}(\Vec{x},x_3,x_4) S^\dagger, ~ a_{\mu}(\Vec{x},x_3,x_4+L) = C a_{\mu}(\Vec{x},x_3,x_4) C^\dagger, \notag \\
    &\lambda (\Vec{x},x_3+L,x_4) = S \lambda(\Vec{x},x_3,x_4) S^\dagger, ~~~~ \lambda (\Vec{x},x_3,x_4+L) = C \lambda (\Vec{x},x_3,x_4) C^\dagger, 
\end{align}
and both gauge fields and fermions get the nonzero KK mass at least of $O(1/NL)$. 
The classical vacuum is obtained by setting $a_3=a_4=0$ and the Polyakov loops are solely determined by the transition function, 
\begin{align}
    P_3 = S,~~~ P_4 = C.
\end{align}
This satisfies $\braket{\operatorname{tr} (P_3^{n_3}P_4^{n_4})} = 0$ for any $(n_3,n_4)\not =(0,0) \bmod N$. 
Thus, we obtain the center-symmetric gapped system, and the $2$d effective theory becomes the discrete $\mathbb{Z}_N$ gauge theory within the perturbation theory.

For the $2$d gapped gauge-Higgs system, it is natural to expect the presence of vortex-like instantons as semiclassical objects. 
To explain its topological stability in our setup, let us compactify $\mathbb{R}^2\times T^2$ to $(T^2)_{12} \times (T^2)_{34}$ with 't Hooft twists $n_{12}=n_{34}=1$ for sufficiently large $L_1, L_2$. 
Then, the topological charge is given by~\cite{vanBaal:1982ag}
\begin{align}
    Q_{\mathrm{top}} \in - \frac{n_{12}n_{34}}{N} + \mathbb{Z} = -\frac{1}{N} +\mathbb{Z}.
\end{align}
Thus, there exists a $2$d localized configuration topologically protected by $Q_{\mathrm{top}}=\pm 1/N$. 
The vortex action $S_v$ is bounded from below by the BPS bound, 
\begin{align}
    S_{v} = \frac{1}{g^2} \int \operatorname{tr}(f\wedge \star f) \geq \frac{8\pi^2}{g^2}|Q_{\mathrm{top}}| = \frac{S_I}{N},
\end{align}
where $S_I=8\pi^2/g^2$ is the $4$d instanton action. The equality holds if and only if the (anti-)self-dual YM equation is satisfied. 

Although it is still an open mathematical problem if the fractional instanton on $\mathbb{R}^2\times T^2$ satisfies the BPS bound, numerical studies~\cite{Gonzalez-Arroyo:1998hjb, Montero:1999by, Montero:2000pb, Wandler:2024hsq} support the existence of the self-dual vortex with $Q_{\mathrm{top}}=\pm 1/N$. 
Moreover, they show that the vortex, or fractional instanton, rotates the phase of Wilson loop by the $\mathbb{Z}_N$ center element when it goes across the loop, which is the characterization of the center vortex.\footnote{
Fractional instantons on $\mathbb{R} \times T^3$ with 't~Hooft twist(s) have been studied numerically in \cite{GarciaPerez:1989gt, GarciaPerez:1992fj, Itou:2018wkm}, and they also satisfy the BPS bound within the numerical accuracy and the lattice discretization error. They play the pivotal role for similar semiclassical approaches in \cite{Yamazaki:2017ulc, Cox:2021vsa}.
See \cite{Fujimori:2019skd,Fujimori:2020zka} for the similar numerical calculations for ${\mathbb C}P^{N-1}$ models on $\mathbb{R} \times S^1$.}
This justifies to perform the semiclassical analysis assuming the self-dual center vortex with $S_v=S_I/N$ and $Q_{\mathrm{top}}=\pm 1/N$.

As the center vortex carries the topological charge $1/N$, it is associated with two fermionic zero modes for $\mathcal{N}=1$ SYM due to the index theorem, $\operatorname{Index}(D) = 2 N Q_{\mathrm{top}} = 2$. 
The presence of the fermionic zero mode tells that the center vortex does not produce the area law of the Wilson loop as its contribution vanishes by the fermion path integral. 
However, it generates the chiral condensate $\langle \operatorname{tr}(\lambda \lambda)\rangle \sim \rme^{-S_I/N}$, because the operator $\operatorname{tr}(\lambda \lambda)$ can absorb the fermionic zero mode of the center vortex when the location of the center vortex coincides with that of the operator insertion. 
Therefore, $\mathcal{N}=1$ SYM on $\mathbb{R}^2\times T^2$ with the 't Hooft flux shows the ``de''-confinement of the $2$d Wilson loops and the discrete chiral symmetry breaking. 
At first glance, the deconfinement of the $2$d Wilson loops may seem to suggest the failure of the adiabatic continuity to the $4$d confinement states, but this is not the case due to the subtle interplay between $(\mathbb{Z}_N^{[1]})_{2\mathrm{d}}$ and $(\mathbb{Z}_{2N})_{\mathrm{chiral}}$~\cite{Tanizaki:2022ngt}. 
The $2$d mixed anomaly between $(\mathbb{Z}_N^{[1]})_{2\mathrm{d}}$ and $(\mathbb{Z}_{2N})_{\mathrm{chiral}}$ comes out of the $4$d 't~Hooft anomaly thanks to the 't~Hooft twisted compactification~\cite{Tanizaki:2022ngt} (see also \cite{Tanizaki:2017qhf, Yamazaki:2017dra, Tanizaki:2017mtm, Dunne:2018hog}), and the gapped vacua must form the projective representation of ($\mathbb{Z}_N^{[1]})_{2\mathrm{d}}\times (\mathbb{Z}_{2N})_{\mathrm{chiral}}$ by the anomaly matching condition. 
This is why we only get the $N$-fold degeneracy of the ground states instead of $N^2$ ones despite the fact that both $(\mathbb{Z}_{N}^{[1]})_{2\mathrm{d}}$ and $(\mathbb{Z}_{2N})_{\mathrm{chiral}}$ are broken.\footnote{The same phenomenon occurs also for the massless charge-$N$ Schwinger model, and all these conclusions can be confirmed by exact calculations~\cite{Anber:2018jdf, Misumi:2019dwq, Armoni:2018bga, Honda:2022edn}.} 

To confirm these claims semiclassically, 
a convenient way to describe the vacuum structure is to compactify $\mathbb{R}^2 \times (T^2)_{34}$ to $\mathbb{R} \times (S^1)_2 \times (T^2)_{34}$ with the size $L_2$, and we set $L = L_{3,4} \ll L_2 \ll \Lambda^{-1}$. 
We then take the Hamiltonian viewpoint with regarding $\mathbb{R}$ as the time direction for the quantization~\cite{Tanizaki:2022ngt, Yamazaki:2017ulc, Cox:2021vsa}. 
As $L=L_{3,4}$ is the shortest length scale, we first minimize along those directions, and the 't Hooft twist $n_{34}=1$ sets $P_3=S$, $P_4=C$ as we have already seen. 
Next, $P_2$ must commute with both $P_3,P_4$ to minimize the classical action along the $2$-$3$ and $2$-$4$ directions, and then we obtain $N$ classical vacua labeled by $P_2 = \rme^{\frac{2 \pi \im m}{N}} 1_{N \times N}$ with $m=0,1,\ldots, N-1$. 
Let us call this state $\ket{m}$:
\begin{align}
    P_2 |m\rangle = \rme^{\frac{2\pi \im m}{N}} |m\rangle. 
\end{align}
When the center vortex does not carry the fermionic zero modes as in the case of pure YM, it produces the transition amplitude, or the tunneling process, from $|m\rangle$ to $|m\pm 1\rangle$ as $\langle m\pm 1, \tau=T| m, \tau=0\rangle \equiv \langle m\pm 1|\exp(-T \hat{H})|m\rangle \sim T \rme^{-S_I/N\pm \im \theta/N}$, and the vacuum degeneracy is lifted. 
For the $\mathcal{N}=1$ SYM, however, the center vortex carries the fermion zero modes, and thus $\langle m\pm 1,T|m,0\rangle=0$, which shows that $|m\rangle$ is the correct eigenstate of the SYM Hamiltonian $\hat{H}_{\mathrm{SYM}}$. 
The insertion of $\operatorname{tr}(\lambda \lambda)$ absorbs two zero modes for $\lambda$ and the center-vortex process contributes to the following matrix element, 
\begin{align}
    \bra{m + 1, T} \operatorname{tr} (\lambda \lambda(\tau)) \ket{m, 0} \sim \frac{1}{L^3} \rme^{-S_I /N + \im \theta /N}\,. \label{eq:amplitude_center-vortex}
\end{align}
Similarly, the anti-vortex contributes to $\bra{m-1,T} \operatorname{tr}(\Bar{\lambda}\Bar{\lambda}(\tau))\ket{m,0}\sim \frac{1}{L^3} \rme^{-S_I/N-\im \theta/N}$. We note that these amplitudes do not have the imaginary-time length $T$ as the overall coefficient because the dominant contribution comes only when the center vortex sits on top of the operator insertion. The prefactor $1/L^3$ is multiplied to match the mass dimension of both sides. 
We can diagonalize the chiral condensate by changing the basis as
\begin{align}
    \widetilde{\ket{k}} := \frac{1}{\sqrt{N}} \sum_{m \in \mathbb{Z}_N} \rme^{-2 \pi \im k m / N} \ket{m},
\end{align}
and then we obtain 
\begin{align}
    \widetilde{\bra{k}} \operatorname{tr} (\lambda \lambda) \widetilde{\ket{k}} \sim \frac{1}{L^3}\rme^{-S_I /N + \im (\theta + 2 \pi k) /N} \sim \Lambda^3 \rme^{\im (\theta+2\pi k)/N}. 
\end{align}
As the price, the Polyakov loop is no longer diagonal, $\widetilde{\bra{k_1}}P_2 \widetilde{\ket{k_2}}=\delta_{k_1, k_2- 1}$. 

Due to the fact that the center vortex rotates the phase of the Wilson loop and also carries the fermionic zero modes, we cannot diagonalize $P_2$ and $\operatorname{tr}(\lambda \lambda)$ simultaneously. 
As a result, the gluino condensate $\operatorname{tr}(\lambda \lambda)$ shows the nontrivial commutation relation with $\operatorname{tr}(P_2)$ on the low-energy Hilbert space $\mathcal{H}_{\mathrm{g.s.}}=\sum_m \mathbb{C}|m\rangle=\sum_k \mathbb{C} \widetilde{|k\rangle}$, 
\begin{align}
    \operatorname{tr} (\lambda \lambda) \operatorname{tr} (P_2) =  \rme^{-2 \pi \im  /N} \operatorname{tr} (P_2) \operatorname{tr} (\lambda \lambda).
\end{align}
This is the semiclassical realization of the 2d mixed anomaly between $(\mathbb{Z}_N^{[1]})_{\mathrm{2d}}$ and $(\mathbb{Z}_{2N}^{[0]})_{\mathrm{chiral}}$ symmetries. 
From this observation, the large Wilson loop becomes the discrete chiral symmetry generator under the renormalization-group flow, and the perimeter law of the $2$d Wilson loop can be understood as the topological property of the spontaneously broken chiral symmetry generator.

Let us pose several puzzles about the 2d center-vortex semiclassics for $\mathcal{N}=1$ SYM:
\begin{itemize}
    \item The derivation of vacuum structure is less straightforward compared with that of $3$d monopole/bion-based semiclassics.
    It would be better to derive the vacuum structure directly from the 2d dilute gas calculation of the center vortices.

    \item Unlike the case of $3$d monopole semiclassics, we lack the analytic formula for the self-dual center vortex, and its fluctuation determinant cannot be calculated. 
    This prohibits us from comparing the results with other analyses quantitatively. 

    \item It is quite counter-intuitive that the Wilson loop in $2$d obeys the perimeter law whereas the vacuum structure is smoothly connected to the $4$d/$3$d confinement vacua. 
    What is the microscopic dynamics realizing the \hl{switching}\footnote{\hl{
    The adiabatic continuity holds between small $L_3$ and large $L_3$ regimes, as shown in Section~\ref{sec:WeakWeakContinuity} by explicitly demonstrating that local observables and vacuum structure are kept under this compactification. 
    However, this might sound to have a conflict with the switching of the Wilson loop behaviors from the area law to the perimeter law, which happens instantaneously upon the $\left(S^1\right)_3$ compactification regardless of how large $L_3$ is.  
    Its consistency with the adiabatic continuity is ensured since this switching occurs for asymptotically large Wilson loops that are much larger than the compactification size $L_3$.
    }} from the 3d area law to the 2d perimeter law? 

    \item In the context of 3d semiclassical analysis, both confinement and chiral symmetry breaking arise from magnetic bions. In the $2$d semiclassics, however, the center vortex does not have this character unlike the BPS and KK monopoles, and it seems that there is no counterpart of magnetic bions. Do magnetic bions play any roles in the $2$d semiclassics?
\end{itemize}
In the following of this paper, we shall address these problems by revising the $2$d center-vortex theory from the viewpoint of the $3$d monopole/bion theory.  

\section{Weak-weak continuity in \texorpdfstring{$\mathcal{N}=1$}{N=1} super-Yang-Mills}
\label{sec:WeakWeakContinuity}

In this section, we aim to understand the 2d center-vortex semiclassics for $\mathcal{N}=1$ SYM theory from the 3d monopole/bion semiclassics by extending the recent achievement that unifies the monopole and center-vortex semiclassics for deformed YM~\cite{Hayashi:2024yjc}. 

\subsection{Monopole-vortex continuity}

We put the $\mathcal{N}=1$ SYM on $\mathbb{R}^2 \times (S^1)_3 \times (S^1)_4 \ni (\bm{x}, x_3,x_4)$ with $x_3 \sim x_3 + L_3$ and $x_4 \sim x_4 + L_4$ using the 't~Hooft twisted boundary condition, where the latter $S^1$ is always small: $L_4 \ll (N\Lambda)^{-1}$. 
We first consider the large-$L_3$ case so that the $3$d monopole/bion effective theory is applicable, and then we gradually make $L_3$ smaller to observe how the monopole/bion semiclassics is connected to the center-vortex effective theory. 

The $3$d monopole/bion effective theory is already explained in Section~\ref{sec:MonopoleSemiclassics}, and we need to identify the role of the 't~Hooft flux along the $3$-$4$ direction in that language. 
From the 3d perspective after the $(S^1)_4$ compactification, the $4$d $1$-form symmetry, $(\mathbb{Z}_N^{[1]})_{4\mathrm{d}}$, becomes 
\begin{equation}
(\mathbb{Z}_N^{[1]})_{4\mathrm{d}}\rightarrow (\mathbb{Z}_N^{[1]})_{3\mathrm{d}}\times (\mathbb{Z}_N^{[0]})_{3\mathrm{d}},  
    \label{eq:3dSymmetry}
\end{equation}
where $(\mathbb{Z}_N^{[1]})_{3\mathrm{d}}$ acts on the spatial Wilson loop, and $(\mathbb{Z}_N^{[0]})_{3\mathrm{d}}$ is the conventional center symmetry, acting on the Polyakov loop $P_4$.
In this 3d language, the 't Hooft flux along the $3$-$4$ direction corresponds to the $(\mathbb{Z}_N^{[0]})_{3\mathrm{d}}$-twisted boundary condition on $\mathbb{R}^2\times (S^1)_3$.
The action of $(\mathbb{Z}_N^{[0]})_{3\mathrm{d}}$ center symmetry on the dual photon and holonomy $(\vec{\sigma}, \vec{\phi})$ is shown in (\ref{eq:3d-center-transf}), and the boundary conditions are\footnote{Note that the fields are subject to the $(\mathbb{Z}_N^{[0]})_{3\mathrm{d}}$ center transformation from $x_3 = L_3$ to $x_3 = 0$.}
\begin{equation}
\vec{\sigma}(\bm{x}, x_3) = P_W \vec{\sigma}(\bm{x}, x_3 + L_3),\quad 
\vec{\phi}(\bm{x}, x_3) = P_W \vec{\phi}(\bm{x}, x_3 + L_3) + 2 \pi \vec{\mu}_1. 
\end{equation}
Since the Weyl vector satisfies $P_W \vec{\rho} = \vec{\rho} - N \vec{\mu}_1$, these boundary conditions can be summarized as
\begin{equation}
\vec{z}(\bm{x}, x_3 + L_3) = P_W^{-1} \vec{z}(\bm{x}, x_3), \label{eq:twisted_BC_bosonic}
\end{equation}
with the complex scalar $\vec{z}$ defined by (\ref{eq:complex_var_Z}).
The same Weyl-permutation-twisted boundary condition is imposed on the Cartan adjoint fermion $\vec{\lambda}$. In summary, the boundary condition for the superfield $\vec{Z}$ reads
\begin{equation}
\vec{Z}(\bm{x}, x_3 + L_3) = P_W^{-1} \vec{Z}(\bm{x}, x_3). \label{eq:twisted_BC_superfield}
\end{equation}

\begin{figure}[t]
\centering
\begin{minipage}{.38\textwidth}
\includegraphics[width= \textwidth]{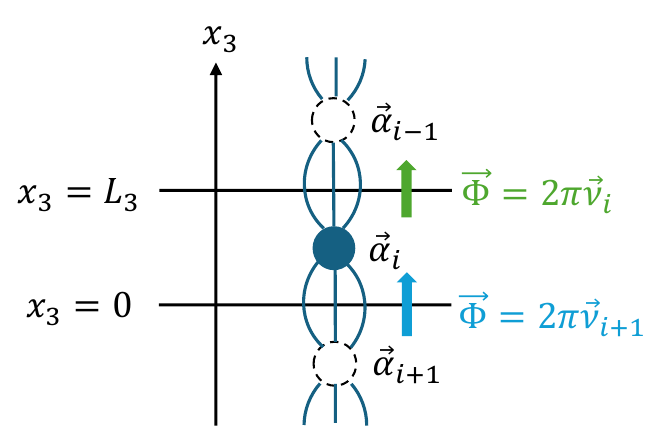}
\end{minipage}\quad
\begin{minipage}{.56\textwidth}
\includegraphics[width= \textwidth]{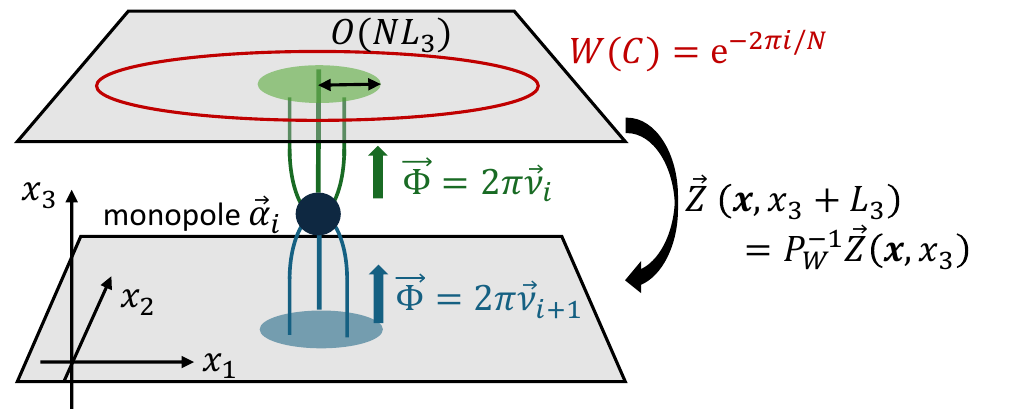}
\end{minipage}
\caption{(\textbf{Left panel}) Mirror-image magnetic monopoles to satisfy the center-twisted boundary condition. 
The magnetic charge $\vec{\alpha}_i$ emits the outgoing magnetic flux $2 \pi \vec{\nu}_i$ and absorbs the incoming magnetic flux $2 \pi \vec{\nu}_{i+1}$ along the $x_3$ direction.
(\textbf{Right panel}) Schematic $3$d view of a BPS/KK monopole in $\mathbb{R}^2 \times S^1$ with the center-twisted boundary condition. The magnetic flux is localized to a size of $O(NL_3)$. This magnetic flux is indeed the center vortex: the Wilson loop acquires the phase $\rme^{-2 \pi \im / N}$ if it surrounds the vortex.}
\label{fig:monopole-vortex-schematic}
\end{figure}

A recent finding in \cite{Hayashi:2024yjc} is that the 2d center-vortex instanton is nothing but the magnetic flux associated with the 3d BPS/KK monopole in $\mathbb{R}^2 \times (S^1)_3$ with $(\mathbb{Z}_N^{[0]})_{3\mathrm{d}}$-twisted boundary condition.
To see this, let us put a BPS or KK monopole $[\mathcal{M}_i(x^* )]\sim \rme^{\vec{\alpha}_i\cdot \vec{z}(x^*)}$, (\ref{eq:monopole-vertex}), which has the magnetic charge $\vec{\alpha}_i$ and is located at $(\bm{0},x_{3,*})$.
To examine the bosonic profile of this monopole, we will analyze the classical equation of motion in the 3d EFT: 
\begin{align}
    \nabla^2  \vec{\phi} &= 2\pi L_4 \vec{\alpha}_i\, \delta^{(2)}(\bm{x})\delta(x_3-x_{3,*}), \notag \\
    \nabla^2  \vec{\sigma} &= - 2 \pi L_4 \vec{\alpha}_i\left( \frac{\theta}{2 \pi} + \frac{4 \pi \im}{g^2} \right)\, \delta^{(2)}(\bm{x})\delta(x_3-x_{3,*}), 
    \label{eq:MagneticCoulombEq}
\end{align}
with the Weyl-permutation-twisted boundary condition (\ref{eq:twisted_BC_bosonic}).
The effect of the boundary condition can be taken into account by extending the range of $x_3$ to $\mathbb{R}$ with the mirror-image method (See the left panel of Figure \ref{fig:monopole-vortex-schematic}).

Let us solve the classical equation of motion with $\vec{\sigma}\to 0$ and $\vec{\phi}\to \vec{\phi}_c$ as $|\bm{x}|\to \infty$. 
The result is given by\footnote{A naive guess from the mirror-image method is
\begin{align}
    \vec{\phi}-\vec{\phi}_c
    = \frac{\im g^2}{4\pi}\left(\vec{\sigma}+\frac{\theta}{2\pi}(\vec{\phi}-\vec{\phi}_c)\right)= \frac{L_4}{2} ``\sum_{n \in \mathbb{Z}} \frac{\vec{\alpha}_{i-n~(\operatorname{mod}N)}}{\sqrt{|\bm{x} - \bm{x}^*|^2+(x_{3}- x_{3,*} - n L_3)^2}}", 
    \label{eq:mirror_image_naive}
\end{align}
For the sake of convergence and satisfying the boundary condition at $|\bm{x}|\to \infty$, we have to change the order of the summation by rewriting $\vec{\alpha}_i=\vec{\nu}_i-\vec{\nu}_{i+1}$ from this naive one.
}
\begin{align}
    &\vec{\phi}-\vec{\phi}_c
    = \frac{\im g^2}{4\pi}\left(\vec{\sigma}+\frac{\theta}{2\pi}(\vec{\phi}-\vec{\phi}_c)\right) \notag\\
    &\,\, = - \frac{L_4}{2}\sum_{\ell=0}^{N-1}\vec{\nu}_{i-\ell} \sum_{k\in \mathbb{Z}}\left(\frac{1}{\sqrt{|\bm{x}|^2+(x_{3,\ell}- NkL_3)^2}}-\frac{1}{\sqrt{|\bm{x}|^2+(x_{3,\ell}-L_3- NkL_3)^2}}\right),  
    \label{eq:SolutionMagneticPotential}
\end{align}
where $x_{3,\ell}=x_3-x_{3,*}-\ell L_3$, and $\vec{\nu}_i$ is the weight vector of the defining representation:
\begin{equation}
    \vec{\nu}_1=\vec{\mu}_1,\,\, \vec{\nu}_2=\vec{\mu}_1-\vec{\alpha}_1,\,\,\ldots, \,\, \vec{\nu}_N=\vec{\mu}_1-\vec{\alpha}_1-\cdots - \vec{\alpha}_{N-1}. \label{eq:weight_vector_def}
\end{equation}
As $|\bm{x}|\to \infty$, $\vec{\sigma}=O(\rme^{-\frac{2\pi}{N L_3}|\bm{x}|})$ and $\vec{\phi}-\vec{\phi}_c=O(\rme^{-\frac{2\pi}{N L_3}|\bm{x}|})$, and thus the magnetic field is exponentially localized at the origin of $\mathbb{R}^2$.

As the $U(1)^{N-1}$ field strength is given by $ \vec{f} = \frac{\im g^2}{4 \pi L_4} \star \left( \diff \vec{\sigma} + \frac{\theta}{2 \pi} \diff \vec{\phi} \right)$ via the Abelian duality, the magnetic flux $\vec{\Phi}$ through the $x_1$-$x_2$ plane can be expressed as\footnote{One can obtain this result by naively swapping the order of the sum and the integral. However, if we do it more correctly by regularizing the $\mathbb{R}^2$ integral into the finite region and then taking the $\mathbb{R}^2$ limit, each term of the summand produces the extra term. Those extra terms cancel out thanks to $\sum_{\ell=0}^{N-1}\vec{\nu}_{\ell} = 0$, and we obtain the same result with the naive manipulation.}
\begin{align}
    \vec{\Phi}(x_3)&=\int_{\mathbb{R}^2} \diff^2\bm{x}\,\frac{\im g^2}{4\pi L_4}\partial_{x_3} \left(\vec{\sigma}(\bm{x},x_3) + \frac{\theta}{2 \pi} \vec{\phi} (\bm{x},x_3) \right)\notag\\
    &=\pi\sum_{\ell=0}^{N-1}\vec{\nu}_{n-\ell} \sum_{k\in \mathbb{Z}}\Big[\mathrm{sign}(x_{3,\ell}-NkL_3) -\mathrm{sign}(x_{3,\ell}-L_3-Nk L_3)\Big].
\end{align}
When $x_3$ is restricted to the physical domain, we have
\begin{equation}
    \vec{\Phi}(x_3)=\left\{
    \begin{array}{cc}
         2\pi \vec{\nu}_{i}& (x_{3,*}<x_3<L_3), \\
         2\pi \vec{\nu}_{i+1}& (0\le x_3<x_{3,*}). 
    \end{array}
    \right.
\end{equation}
Thus, the magnetic field out of the BPS/KK monopole takes the vortex-like profile shown in Figure~\ref{fig:monopole-vortex-schematic}: It localizes in the $\mathbb{R}^2$ direction and forms the vortex along $(S^1)_3$ with the outgoing magnetic flux $2 \pi \vec{\nu}_i$ and the incoming magnetic flux $2 \pi \vec{\nu}_{i+1}$.

Then, in the presence of the BPS/KK monopole, the 2d Wilson loop $W(C)$ on the $x_1$-$x_2$ plane has the phase
\begin{align}
    W(C) &= \frac{1}{N}\sum_{\ell' =1}^N \exp\left({\im \vec{\nu}_{\ell'} \cdot \int_C \vec{a}}\right) \notag \\
    &= \left\{
    \begin{array}{cl}
       \rme^{-2\pi \im/N}  & (\text{when $C$ surrounds the localized flux}), \\
       1  & (\text{when $C$ does not surround the localized flux}),
    \end{array}
    \right. \label{eq:center-magnetic-flux}
\end{align}
which is the characterization of the center vortex.
As we advocated, the $2$d center vortex is obtained as the magnetic flux out of the BPS/KK monopole on $\mathbb{R}^2\times (S^1)_3$ with the $(\mathbb{Z}_N^{[0]})_{3\mathrm{d}}$-twisted boundary condition. 
This is the essential ingredient to understand the $2$d center-vortex semiclassics from the viewpoint of $3$d monopole/bion semiclassics. 

We note that the BPS and KK monopoles are permuted under the $(\mathbb{Z}_N^{[0]})_{3\mathrm{d}}$ center symmetry. 
Therefore, the distinction between the BPS/KK monopoles disappears in this setup due to the twisted boundary condition by extending the internal moduli of the center vortex as $x_3 \in [0,NL_3)$: All the BPS and KK monopoles are connected as the single chain via the center-vortex magnetic flux in the extended moduli space. 
This explains the uniqueness of the $2$d center vortex despite the presence of $N$ distinct fundamental monopoles in $3$d semiclassics. 

As emphasized in \cite{Hayashi:2024yjc}, this picture provides the weak-coupling realization of the scenario that the monopole serves as the kink of the center-vortex network~\cite{DelDebbio:1997ke, Ambjorn:1999ym, deForcrand:2000pg}. 
In the $4$d center-vortex model, the presence of monopoles in the vortex network makes the vortex surface nonorientable, which plays the pivotal role to have nonzero instanton numbers~\cite{Engelhardt:1999xw, Reinhardt:2001kf, Cornwall:1999xw}. 
Such nonoerientable vortex surfaces give a model explaining asymptotic string tensions~\cite{Oxman:2018dzp,  Junior:2022bol}. 
The importance of the presence of monopoles in addition to center vortices for confinement is also recently pointed out in \cite{Nguyen:2024ikq}.

\subsection{Weak-weak continuity of superpotential and gluino condensate}

We next investigate the continuity of the vacuum structure between the $3$d monopole/bion regime and the $2$d center-vortex regime. 
For this purpose, let us discuss how the vacuum configuration is chosen for large $L_3$ and small $L_3$ regimes. 

When $L_3$ is large enough, the effect of the twisted boundary condition is negligible and the vacuum is determined by the minima of the bosonic potential. 
The bion-induced bosonic potential is given by \eqref{eq:BionPotential}, and we find the confining chiral-broken vacua, $\vec{z}=\frac{2\pi k\vec{\rho}}{N}\im$ for $k=0,1,\ldots, N-1$. 

When $L_3$ becomes sufficiently small, we need to minimize the kinetic term along the $(S^1)_3$ direction rather than the potential, and thus we must find the constant mode that satisfies the boundary condition. 
The constant mode with the center-twisted boundary condition gives 
\begin{equation}
    \vec{z}=P_W^{-1}\vec{z}. 
\end{equation}
Due to the periodicity of dual photons, this equation has to be solved up to the identification, $\vec{z}\sim \vec{z}+2\pi \im \vec{\mu}_n$ ($n-1,\ldots, N-1$). 
There are only $N$ distinct constant configurations, 
\begin{equation}
    \vec{z}=\frac{2\pi k \vec{\rho}}{N}\im, \quad (k=0,1,\ldots, N-1),
\end{equation}
which satisfies the requirement. These are exactly the same as the vacuum configurations for large $L_3$ limit. This justifies the weak-weak continuity of the vacuum structure between the large and small $L_3$ limits.

Let us reconsider the $2$d semiclassical analysis using the $3$d monopole/bion semiclassics. 
The $2$d effective theory is completely gapped at the perturbative level due to the twisted boundary condition: all adjoint gluon and gluino acquire the $O(1/NL_3)$ KK mass. There are no gapless fluctuations but we still need to take into account the $N$ distinct vacua. 
In Section~\ref{sec:VortexSemiclassics}, we computed the transition amplitude instead of the partition function to circumvent this issue, but we here take a more field-theoretic approach to compute the superpotential.  
We use the technique of ``integrating in'' the auxiliary field $\Sigma$ to include the $\mathbb{Z}_N$ discrete low-energy mode. 

Let us denote  
\begin{align} 
Z_1 := \vec{\alpha}_1 \cdot \vec{Z}, \cdots, Z_{N-1} := \vec{\alpha}_{N-1} \cdot \vec{Z}, 
\end{align}
then $\vec{\alpha}_N\cdot \vec{Z}=-Z_1-\cdots-Z_{N-1}$. 
We would like to treat 
\begin{equation}
    Z_N:=\vec{\alpha}_N\cdot \vec{Z}\,,
\end{equation} 
on the equal footing with $Z_1,\ldots, Z_{N-1}$ 
by regarding them as independent fields, 
and then the constraint $Z_1 + \cdots + Z_N = 0~(\operatorname{mod} 2 \pi \im )$\footnote{The $2 \pi \im$ periodicity of $Z_i$ comes out of the weight vector periodicity of the dual photon $\vec{\sigma} \sim \vec{\sigma} + 2 \pi \vec{\mu}_k~~(k=1,\cdots,N-1)$. The effective potential seems to violate the periodicity of the holonomy field, but we should notice that this effective description breaks down at the boundary of the Weyl chamber due to the presence of extra massless modes. Therefore, we only need to take into account the periodicity of dual photons within this Abelianized effective theory.} should appear as the equation of motion. 
This can be achieved by rewriting the superpotential~\eqref{eq:AffineTodaW} in terms of of $(Z_1, \cdots, Z_N)$ and the auxiliary field $\Sigma$,
\begin{align} 
W(Z_1, \cdots, Z_N, \Sigma)  = L_4 \Lambda^3 \rme^{ \frac{\im \theta}{N}} \sum_{i=1}^{N} \rme^{Z_i} + \Sigma \left( \rme^{Z_1 + \cdots + Z_N} -1 \right).
\end{align}
When we solve the equation of motion of $\Sigma$, we obtain the Affine Toda superpotential~\eqref{eq:AffineTodaW}. 

Since the center symmetry acts as $Z_i\mapsto Z_{i+1}$, the center-twisted boundary condition~\eqref{eq:twisted_BC_superfield} becomes 
\begin{align} 
    Z_i (\bm x, x_3+L_3) = Z_{i+1} (\bm x, x_3). 
\end{align}
This twisted boundary condition gives the KK mass except for the diagonal $U(1)$ mode, $\rme^{Z} := \rme^{Z_1} = \cdots = \rme^{Z_N}$.
By integrating out the higher KK modes, the effective superpotential becomes\footnote{As a side remark, we could introduce the auxiliary field in a different way as $-\tilde{\Sigma}(Z_1+\cdots+Z_N)$ in the superpotential neglecting the subtlety related to the periodicity of $Z_i$. 
Integrating out $(Z_1,\cdots,Z_N)$, we get the effective superpotential in terms of the auxiliary field $\tilde{\Sigma}$, which reads
$W (\tilde{\Sigma}) = N \tilde{\Sigma}  - N \tilde{\Sigma} \log \left( \frac{\tilde{\Sigma}}{L_4 \Lambda^3 \rme^{ \frac{\im \theta}{N}}} \right)$. 
This is nothing but the Veneziano-Yankielowicz superpotential~\cite{Veneziano:1982ah}. Since $\tilde{\Sigma}$ is neutral under the center symmetry, we can also discuss the weak-weak continuity based on this effective superpotential. 
}
\begin{align} 
W(Z, \Sigma)  = N L_4 \Lambda^3 \rme^{ \frac{\im \theta}{N}} \rme^{Z} + \Sigma \left( \rme^{N Z} -1 \right).
\end{align}
This superpotential governs the 2d effective theory on $\mathbb{R}^2$.
The equations of motion gives
\begin{align} 
\rme^{N Z} =1 ,~~~\Sigma = - L_4 \Lambda^3 \rme^{ \frac{\im \theta}{N}} \rme^{Z}\,.
\end{align}
This leads us to the same $N$ vacua as found in the 3d monopole semiclassics. 
We note that the gluino condensate $\langle \operatorname{tr}(\lambda \lambda)\rangle$ can be obtained by taking the derivative of $W$ in terms of the (holomorphic) gauge coupling, and we can find 
\begin{equation}
    \langle \operatorname{tr}(\lambda\lambda)\rangle_k = 16\pi^2 \Lambda^3 \rme^{\im (\theta + 2\pi k)/N}, 
\end{equation}
with $k=0,1,\ldots N-1$~\cite{Davies:1999uw, Davies:2000nw}, which is independent of $L_3$ and $L_4$.\footnote{Recently, the direct semiclassical computation of the gluino condensate on $T^4$ with the 't~Hooft twist is carried out in Refs.~\cite{Anber:2022qsz, Anber:2024mco}, and the result is consistent with the weak-coupling instanton calculus. }  
We would like to point out that the numerical coefficient for the gluino condensate was not determined within the $2$d center-vortex semiclassics in Ref.~\cite{Tanizaki:2022ngt}, and this becomes possible by revising it from the 3d monopole/bion semiclassical description.

\subsection{Mass deformation and phase diagram}
\label{sec:MassDeformation}

\begin{figure}[t]
\centering
\includegraphics[width = 0.50 \linewidth]{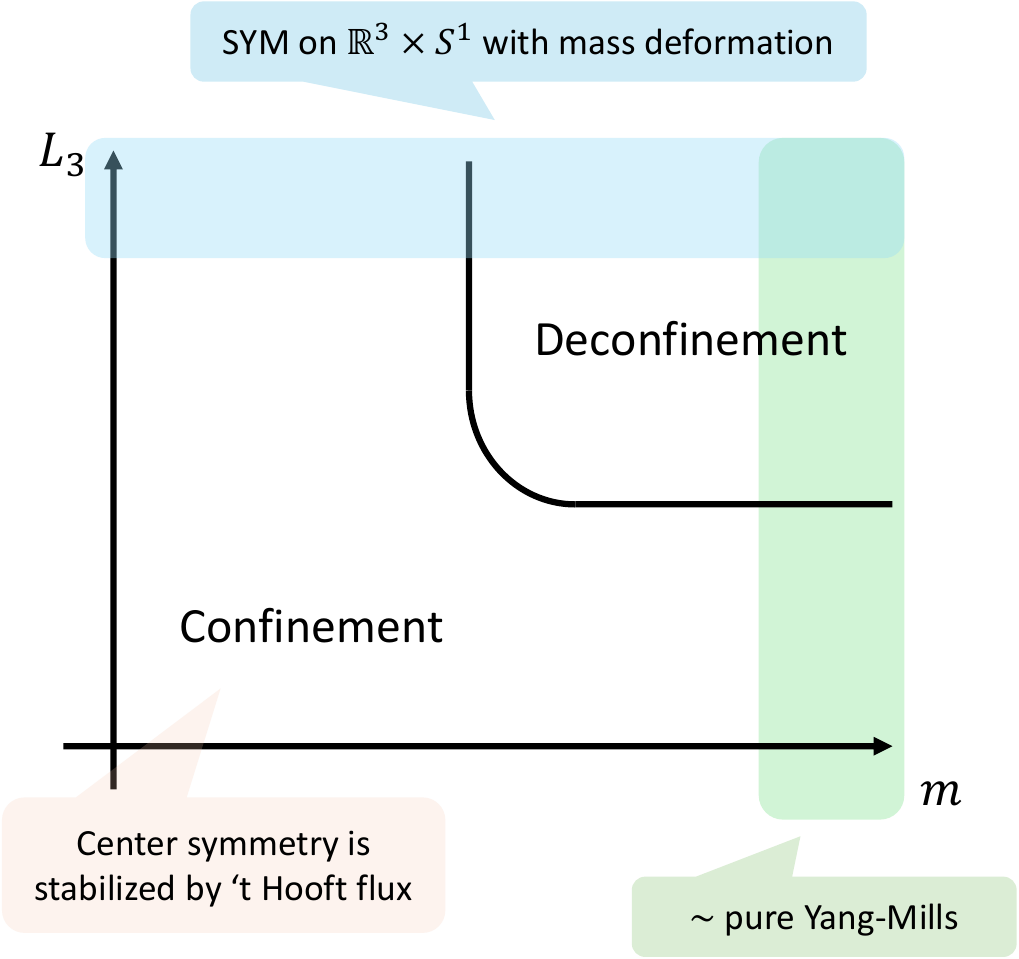}
\caption{Expected behaviors of the phase diagram on the $(L_3, m)$-plane with $N L_4 \Lambda \ll 1$. 
For large $m$ (green), we have a pure Yang-Mills theory on $\mathbb{R}^2 \times S^1 \times S^1$ with the 't Hooft flux. 
For $NL_3 \Lambda \ll 1$, the theory is confined due to the 't Hooft flux, but when $NL_3 \Lambda \gg 1$, the system becomes the high-temperature Yang-Mills theory  and it is in the deconfinement phase.
The large $L_3$ region (blue) should be described as the SYM on $\mathbb{R}^3 \times S^1$ with the mass deformation, and there is also a deconfinement transition at $m_*\sim \Lambda^3 L_4^2$. 
The minimal scenario is to connect these deconfinement transitions by a single transition line. 
}
\label{fig:schematic_phase_diagram}
\end{figure}

If we add the mass term for the adjoint fermion, 
\begin{align} 
\frac{m}{16\pi^2} \left( \operatorname{tr} (\lambda \lambda) + \operatorname{tr} (\Bar{\lambda} \Bar{\lambda}) \right)\,,
\end{align}
the supersymmetry and the discrete chiral symmetry are explicitly broken. 
We assume that $m$ is small, and then the leading correction comes from the fact that the fermionic zero modes of the monopole vertex can be absorbed by the mass term. The $3$d bosonic effective Lagrangian becomes~\cite{Poppitz:2012nz, Anber:2014lba} 
\begin{equation}
    \mathcal{L}_{\mathrm{eff},m}=\frac{g^2}{16\pi^2 L_4}|\diff \vec{z}|^2+V_0\sum_{i=1}^{N} |\rme^{\vec{\alpha}_i\cdot \vec{z}}-\rme^{\vec{\alpha}_{i+1}\cdot \vec{z}}|^2-\frac{m\Lambda^3 L_4}{N} \sum_{i=1}^{N} (\rme^{\vec{\alpha}_i\cdot \vec{z}+\frac{\im \theta}{N}}+\rme^{\vec{\alpha}_i\cdot \vec{z}^*-\frac{\im \theta}{N}}),   
\end{equation}
where the last term describes the contribution of the monopole. 
Both the monopole and bion potential gives the mass to the dual photon and the Wilson loop obeys the area law. However, the monopole term tends to violate the $0$-form center symmetry and leads to the deconfinement transition as $m$ increases. On the $\mathbb{R}^3$ limit, i.e. $L_3\to \infty$, the transition point is roughly given by $m_*\sim \Lambda^3 L_4^2$.

When $m<m_*$, the center-symmetric configurations $\vec{z}=\frac{2\pi k \vec{\rho}}{N}\im$ are the local minima even with the monopole perturbation. 
The role of monopole is to split the $N$ degenerate vacua, and the 2d vacuum energy density, $\int_0^{L_3}\diff x_3 \mathcal{L}_{\mathrm{eff},m}$, reads
\begin{align} 
E_k(\theta) &= - L_3 L_4 m \left(\frac{1}{16\pi^2}\braket{\operatorname{tr} (\lambda \lambda)}_k + c.c.\right) 
    = - 2 L_3 L_4 m \Lambda^3 \cos \left( \frac{\theta + 2 \pi k}{N}\right).
\end{align}
We then obtain the multi-branch structure of the confinement vacua with the level crossing at $\theta=\pi$. Due to the explicit violation of the chiral symmetry breaking, the $2$d Wilson loop also obeys the area law, and then the weak-weak continuity between the $3$d and $2$d semiclassics is valid for the small-$m$ perturbation. 

The above analysis is based on the $3$d monopole/bion semiclassics, and thus we have implicitly assumed that $L_3\gg L_4$. 
When both $L_3$ and $L_4$ are sufficiently small compared with $(N\Lambda)^{-1}$, the 't Hooft twisted compactification stabilizes the center symmetry at the tree level.
This suggests that the system is in the confined phase at any values of $m$ if $L_3$ and $L_4$ are both small. 
The phase diagram on the $(L_3, m)$ plane is sketched in Figure~\ref{fig:schematic_phase_diagram}.

\subsection{Generalization to QCD(adj)}

So far, we have seen that the adiabatic continuity holds between $\mathbb{R}^3 \times S^1$ and $\mathbb{R}^2 \times T^2$ setups. 
The vacua are preserved under the twisted compactification, and the 3d monopole-instanton becomes the 2d center-vortex-instanton in the 2d theory.
It is straightforward to generalize the above argument in the adjoint QCD [QCD(adj)] with $n_f$ quarks.

The QCD(adj) consists of gluons and $n_f$ adjoint Weyl fermions.
In particular, the one-flavor massless QCD(adj) is nothing but the $\mathcal{N}=1$ SYM.
The main difference is the holonomy degrees of freedom.
In the $\mathcal{N}=1$ SYM, the 3d effective theory includes the holonomy variable $\vec{\phi}$.
For $n_f \geq 2$, the perturbative GPY potential fixes the holonomy to the center-symmetric value, and only the dual photon $\vec{\sigma}$ is the low-energy degrees of freedom.
Without the power of the supersymmetry, the magnetic bions such as BPS-$\overline{\mathrm{KK}}$ and their anti-bions still induce the following potential \cite{Unsal:2007vu, Unsal:2008ch}
\begin{align} 
V(\vec{\sigma}) \sim \Lambda^3 (\Lambda L_4)^{3-\frac{4}{3}(n_f-1)} \sum_{i=1}^N \left| \rme^{\im \vec{\alpha}_i \cdot \vec{\sigma}} - \rme^{\im \vec{\alpha}_{i-1} \cdot \vec{\sigma}}  \right|^2,
\end{align}
which gives the same vacua:
\begin{align} 
\vec{\sigma} =  \frac{2 \pi k \vec{\rho}}{N}~~(k=0,\cdots,N-1),
\end{align}
and they are invariant under the $\left( \mathbb{Z}_N^{[0]} \right)_{\mathrm{3d}}$-twisted boundary condition.
This leads us to the 3d-2d adiabatic continuity: these vacua persist under the reduction from the 3d description to the 2d one.

During the twisted compactification, all adjoint fields get $O(1/NL_3)$ Kalzua-Klein mass and decouple.
Thus, the 2d theory only has the $\mathbb{Z}_N$ degrees of freedom of the dual photon, representing the $N$ degenerate vacua, just as in the $\mathcal{N}=1$ SYM.


\section{Double confining string: from 3d area law to 2d perimeter law}
\label{sec:DoubleStringPicture}

Because of the bion mechanism, the Wilson loop obeys the area law in the 3d semiclassics.
On the other hand, the $2$d Wilson loop on $\mathbb{R}^2\times T^2$ with 't~Hooft flux obeys the perimeter law, and it acts as the generator of the spontaneously broken discrete chiral symmetry. as explained in Section~\ref{sec:VortexSemiclassics}.
In this section, we elucidate the microscopic dynamics behind this curious behavior using the 3d monopole/bion semiclassics.

\subsection{Demonstration for the \texorpdfstring{$SU(2)$}{SU(2)} case}

To simplify and focus on the essence, let us first look at the case of $SU(2)$. 
Throughout this section, we assume $L_3 \gg L_4$ so that we can apply the 3d monopole/bion effective theory in $\mathbb{R}^2 \times S^1$.
The field contents of 3d effective theory consist of the dual photon $\sigma$ with the periodicity $\sigma \sim \sigma + 2 \pi$, the holonomy $\phi$, and Cartan Weyl fermion.
With the notation of the previous sections, we can identify the bosonic fields as $\sigma = \vec{\alpha}_1  \cdot \vec{\sigma}$, $\phi = \vec{\alpha}_1  \cdot \vec{\phi}$.

\subsubsection{Wilson loop in the 3d monopole semiclassics}

In the following of this section, we neglect the holonomy field $\vec{\phi}$ by always setting it to the center-symmetric point, and the effective bosonic action in Section~\ref{sec:MonopoleSemiclassics} becomes
\begin{equation}
    S=\frac{g^2}{32\pi^2 L_4}|\diff \sigma|^2+4 V_0 [1-\cos(2\sigma)],  
\end{equation}
where $V_0$ is the bion amplitude. 
Our argument below can easily be generalized to QCD(adj). Recall that the only difference between $\mathcal{N}=1$ SYM and QCD(adj) is the absence of the holonomy variable as it is fixed to the center-symmetric value by the perturbative GPY potential for QCD(adj).

In terms of the dual photon, the Wilson loop $W(C)$ on $\mathbb{R}^3$ is defined as the defect operator that determines the boundary condition of the dual photon $\sigma$ near the loop $C$: For any small loops $C'$ linking to $C$, the dual photon must have a nontrivial monodoromy, 
\begin{align}
    \int_{C'} \diff \sigma = 2 \pi.
\end{align}
We may express this defect as the violation of the magnetic Bianchi identity, $\diff(\diff \sigma)=2\pi \delta[C]$. 

The cos-type nonperturbative potential tries setting $\sigma$ to its minima as much as possible, while the Wilson loop requires that $\sigma$ should go from $0$ to $2\pi$ around the loop. 
To satisfy these requirements, the region where $\sigma$ moves is restricted around the minimal surface attached to the loop $C$, and the confining flux tube is formed:
\begin{equation}
    \langle W(C)\rangle \sim \exp(-T_{\mathrm{conf}}\, \mathrm{Area}(C)),  
\end{equation}
where $T_{\mathrm{conf}}$ denotes the string tension. 
If we look more closely at this flux tube, it turns out to consist of two kinks for $\mathcal{N}=1$ $SU(2)$ SYM theory. 
As the magnetic bion has the magnetic charge $2$, the bion potential takes the form of $-\cos(2\sigma)$, and we have two vacua $\sigma =0,\pi$ $\bmod\, 2\pi$. 
Thus, the kink solution for $\sigma$ going from $0$ to $2\pi$ consists of two minimal kinks, one of which goes from $0$ to $\pi$ and the other goes from $\pi$ to $2\pi$, and we call it the ``double string picture'' following \cite{Anber:2015kea, Bub:2020mff}. 
If we denote the tension of the minimal kink configuration as $T_{\mathrm{kink}}$, we may expect that\footnote{According to the numerical work of Ref.~\cite{Bub:2020mff}, the double-string picture for $\mathcal{N}=1$ $SU(N)$ SYM is valid for $N\le 5$, but the double string collapses to the single string for the fundamental loop of $N\ge 6$.} 
\begin{equation}
    T_{\mathrm{conf}}\approx 2T_{\mathrm{kink}}. 
\end{equation}
Each minimal kink describes the domain wall associated with the discrete chiral symmetry breaking, and thus the confining string consists of two chiral domain walls.

\subsubsection{Wilson loop in \texorpdfstring{${\mathbb R}^2 \times S^1$}{R2xS1}: perimeter law from double string picture}

Now, we consider the behavior of the Wilson loop in the $(\mathbb{Z}_2^{[0]})_{3\mathrm{d}}$-twisted $\mathbb{R}^2 \times (S^1)_3$ setup. 
For $SU(2)$, the Weyl permutation acts as
\begin{align}
    \sigma \rightarrow - \sigma ~~(\operatorname{mod} 2\pi). \label{eq:twisted_BC_SU2}
\end{align}
For the holonomy field, $\phi = \pi$ is not only the minimum of the neutral-bion potential but also consistent with the twisted boundary condition, so we may neglect its fluctuation in the discussion below.
The dual photon feels the magnetic-bion potential, $-\cos (2\sigma)$.
Let us put the Wilson loop $W(C)$ so that $C$ lives in a constant-$x_3$ $\mathbb{R}^2$ plane.

\begin{figure}[t]
\centering
\begin{minipage}{.47\textwidth}
\includegraphics[width= \textwidth]{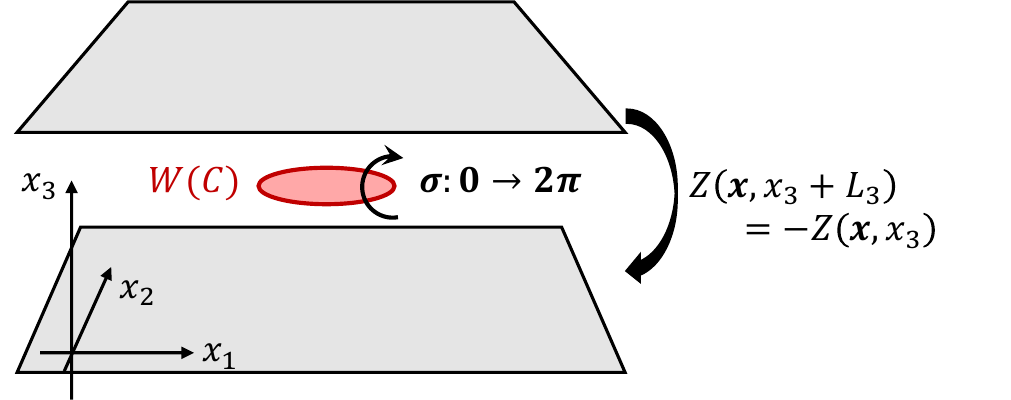}
\end{minipage}\quad
\begin{minipage}{.47 \textwidth}
\includegraphics[width= \textwidth]{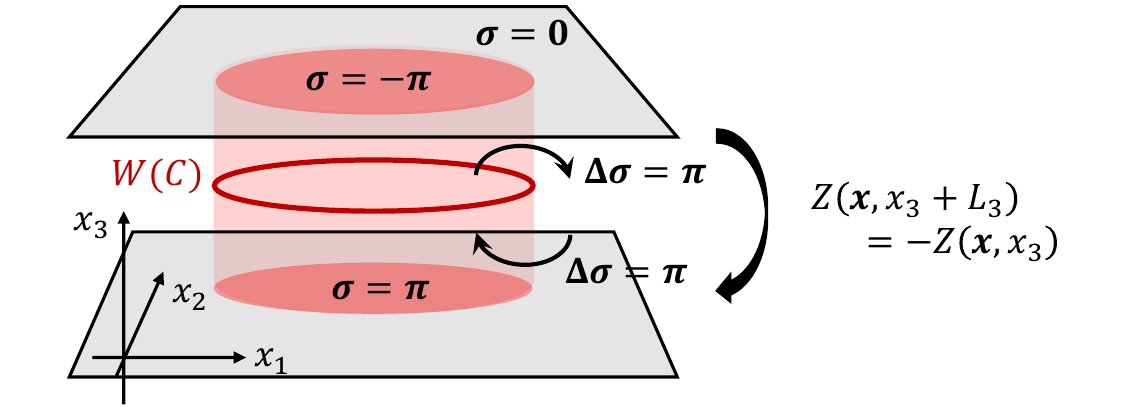}
\end{minipage}
\caption{Two dominant dual-photon configurations for the Wilson loop in the center-twisted $\mathbb{R}^2\times (S^1)_3$. (\textbf{Left panel}) Confinement flux tube forms the minimal-area surface of $C$ in the constant $x_3$ plane. For $SU(2)$, this flux tube has the double kink structure, which is called the double string picture. This contribution gives the dominant contribution when $C$ is not so large. 
(\textbf{Right panel}) The minimal-area surface of $C$ wraps around the $(S^1)_3$ direction: One part of the double string goes upward and the other goes downward out of $C$, and it forms the domain wall for the spontaneously broken chiral symmetry. This configuration is possible due to the twisted boundary condition. When $|C| \gg L_3$, this gives the dominant contribution.}
\label{fig:double-string}
\end{figure}

As illustrated in Figure \ref{fig:double-string}, there are two dominant configurations  of the dual photon that contributes to $\langle W(C)\rangle$ in the center-twisted $\mathbb{R}^2\times (S^1)_3$. 
One is the standard minimal-surface configuration as in the case of $\mathbb{R}^3$, shown in the left panel of Figure \ref{fig:double-string}. 

The other contribution is specific to the compactified geometry: We consider the chiral domain wall attached to $W(C)$ and extended along the $x_3$ direction as shown in the right panel of Figure \ref{fig:double-string}. 
Let us check if this configuration is really consistent with the boundary condition. 
To have the nontrivial winding number of $\sigma$ around the Wilson loop, we have to attach the domain walls with the opposite orientation above and below $W(C)$. 
Since the $x_3$ direction is twisted by the center symmetry, these chiral domain walls with opposite orientation are smoothly glued at $x_3=0\sim L_3$.\footnote{There is another insightful way to understand that the domain wall along the $x_3$ direction is the valid configuration. 
By extending the range of $x_3$ from $\mathbb{R}/L_3\mathbb{Z}$ to $\mathbb{R}$, we need to put the mirror image of the Wilson loop repeatedly as $x_3\mapsto x_3+L_3$ with flipping its charge. The above discussion shows that we can consider the chiral domain wall extended along the $x_3$ direction connecting these Wilson loops without costing much energy. 
This is indeed the case as the Wilson loops are deconfined on the chiral domain wall even though they are confined on the $\mathbb{R}^3$ bulk~\cite{Anber:2015kea}. From the modern view, different chiral-broken vacua are distinct as the symmetry-protected topological states with the center symmetry, and thus the deconfinement on the chiral domain wall is mandatory to satisfy the anomaly inflow from the bulk~\cite{Gaiotto:2017yup, Komargodski:2017smk, Cox:2019aji, Cox:2024tgo}. }
This contribution to $\langle W(C)\rangle$ can be evaluated by the wall tension $T_\mathrm{kink}$ and the area of the wall $L_3\times \mathrm{Length}(C)$ as $\exp(-T_{\mathrm{kink}}\, L_3\, \mathrm{Length}(C))$. 
The dual photon takes $\sigma =0 $ outside the wall and $\sigma=\pi$ inside the wall, and they have the same energy density as these are related by the broken chiral symmetry. 

To sum up, the Wilson loop average gets the contributions from both of these configurations, and which of these configurations dominates depends on the size of the loop.
Schematically, we have
\begin{align}
    W(C) &\sim \rme^{-  T_{\mathrm{conf}}\,\mathrm{Area}(C)} + \rme^{- T_{\mathrm{kink}}\, L_3\, \mathrm{Length}(C)} \notag \\
    &\sim \left\{
    \begin{array}{cl}
         \rme^{-  T_{\mathrm{conf}}\, \mathrm{Area}(C)} & \quad \text{if $C$ is not that large}, \\
        \rme^{- T_{\mathrm{kink}}\, L_3 \mathrm{Length}(C)} & \quad \text{if $C$ is sufficiently large}, 
    \end{array}
    \right.
\end{align}
where $T_{\mathrm{kink}}$ is the kink tension, and $T_{\mathrm{conf}}$ is the string tension. 
If the size of $C$ is not so large compared with $L_3$, the standard area law appears. 
When the size of $C$ becomes large enough compared to $L_3$, the domain wall extended along $(S^1)_3$ dominates and we obtain the perimeter law. 
The \hl{switching scale} between these behaviors can be estimated as $T_{\mathrm{conf}}\, \mathrm{Area}(C_*) = T_{\mathrm{kink}}\, L_3\, \mathrm{Length}(C_*)$; when we use the double-string scenario with $T_{\mathrm{conf}}\approx 2T_{\mathrm{kink}}$, the \hl{switching size} becomes $L_*\approx 2 L_3$ for a square loop of size $L\times L$.
This is the microscopic explanation for why the Wilson loops in the $2$d center-vortex semiclassics shows the perimeter law and generates the spontaneously-broken chiral transformation.

As discussed in Section~\ref{sec:MassDeformation}, the gluino mass term explicitly breaks the discrete chiral symmetry, and the $N$-fold degeneracy is resolved at generic values of the vacuum angle $\theta$.
Then, the energy densities inside and outside the domain wall associated with the Wilson loop become different, and the 2d Wilson loops obey the area law in the presence of mass deformation.
At $\theta=\pi$, there is the $2$-fold ground-state degeneracy due to the spontaneously broken $CP$ symmetry, and the fundamental Wilson loop in the specific orientation follows the perimeter law as its associated domain wall relates the $CP$-broken vacua.

\subsection{Generalization to \texorpdfstring{$SU(N)$}{SU(N)}}

We now extend our previous discussion for the $SU(2)$ gauge group to the $SU(N)$ gauge group. We neglect the fluctuation of the holonomy by setting $\vec{\phi}=\vec{\phi}_c$, and 
the effective action for the dual photons with the bion-induced potential is given by
\begin{align}
    S_{\mathrm{3d, bosonic}} = \frac{g^2}{16 \pi^2 L_4} \left| \diff \vec{\sigma}  \right|^2 + V_0 \sum_{i=1}^N \left| \rme^{\im \vec{\alpha}_i \cdot \vec{\sigma}} - \rme^{\im \vec{\alpha}_{i-1} \cdot \vec{\sigma}}  \right|^2. 
\end{align}
As shown in \eqref{eq:vacua_in_Z}, this potential has the $N$ vacua $\vec{\sigma} = \frac{2 \pi k }{N}\vec{\rho}$ with $k=0,1,\ldots, N-1$, and the discrete chiral symmetry is spontaneously broken.

Let us consider the behavior of the fundamental Wilson loop. In the Abelianized regime, the fundamental Wilson loop is represented as 
\begin{equation}
    \operatorname{tr}[W(C)]=\sum_{i=1}^{N}\exp\left(\im \oint_{C}\vec{\nu}_i\cdot \vec{a}\right), 
\end{equation}
where $\vec{a}$ is the Cartan gluons and $\vec{\nu}_i=\vec{\mu}_1-(\vec{\alpha}_1+\cdots \vec{\alpha}_{i-1})$. 
In the 3d semiclassics, each component, $\exp(\im \oint_C\vec{\nu}_i\cdot \vec{a})$, of the fundamental Wilson loop can be translated as the defect operator that imposes the nontrivial winding number, 
\begin{align}
    \int_{C'} \diff \vec{\sigma} = 2 \pi \vec{\nu}_i,
\end{align}
for any small loops $C'$ linking to $C$; equivalently, one may express this condition as 
\begin{align}
    \diff(\diff \vec{\sigma})=2\pi \vec{\nu}_i\, \delta[C], 
\end{align} 
where $\delta[C]$ is the delta-functional $2$-form for the closed loop $C$. 
Since all these $N$ components are related by the unbroken $(\mathbb{Z}_N^{[0]})_{3\mathrm{d}}$ center symmetry, they show exactly the same behavior and we focus on the case of $\vec{\nu}_1=\vec{\mu}_1$. 

\begin{figure}[t]
\centering
\includegraphics[width = 0.75 \linewidth]{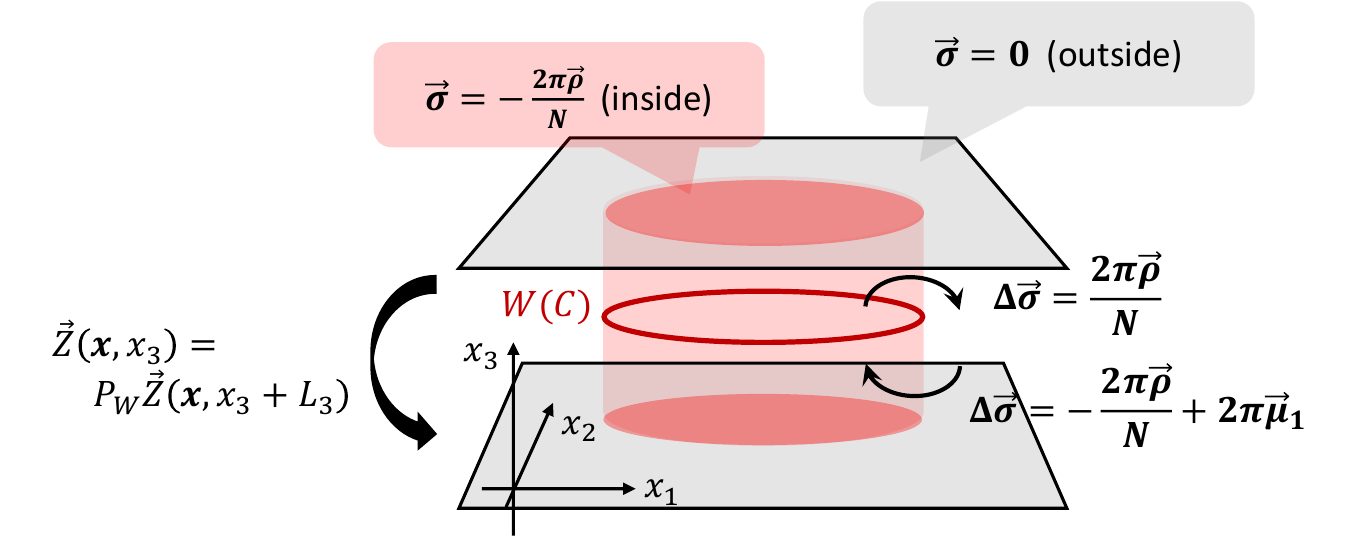}
\caption{Dominant configuration for $|C| \gg L_3$. One of the kinks from the Wilson loop extends upward, while the other extends downward. This configuration is possible due to the twisted boundary condition: $(\Delta \vec{\sigma})_{\mathrm{lower~kink}}=- P_W (\Delta \vec{\sigma})_{\mathrm{upper~kink}}$, and explains the 2d perimeter law.}
\label{fig:double-string-su_n}
\end{figure}

On $\mathbb{R}^3$, the Wilson loop shows the area law, $\langle W(C)\rangle \sim \exp[-T_{\mathrm{conf}}\, \mathrm{Area}(C)]$. The magnetic bion potential tries to set the field configuration to one of its vacuum, say $\vec{\sigma}=\vec{0}$, as much as possible, but the Wilson loop imposes that $\vec{\sigma}$ should goes from $\vec{0}$ to $2\pi \vec{\mu}_1$ around the loop. This causes the formation of the kink on which $\vec{\sigma}$ rapidly changes from $\vec{0}$ to $2\pi \vec{\mu}_1$ and $\vec{\sigma}$ stays almost constant everywhere away from the kink, which is the valid configuration due to the gauge redundancy, $\vec{\sigma}\sim \vec{\sigma}+2\pi \vec{\mu}_1$. 
The energy density of this kink gives the confining string tension, $T_{\mathrm{conf}}$. 

On $\mathbb{R}^2\times (S^1)_3$ with the twisted boundary condition, there is the other configuration that also contributes to $\langle W(C)\rangle$. 
The twisted boundary condition~\eqref{eq:twisted_BC_bosonic} requires $\vec{\sigma}(\bm{x},x_3+L_3)= P_W^{-1} \vec{\sigma}(\bm{x},x_3)$, and an analog of the right panel of Figure \ref{fig:double-string} can be generalized to the $SU(N)$ case as follows (its details are illustrated in Figure~\ref{fig:double-string-su_n}): The kink extended along the $x_3$ direction is created by the Wilson loop, and the dual photon $\vec{\sigma}$ take the following expectation values in the bulk region, 
\begin{equation}
\vec{\sigma}(\bm{x},x_3) = \begin{cases}
    \vec{0} ~~&(\text{$\bm{x}$ is outside of the wall attached to $C$}), \\
    -\frac{2 \pi}{N}\vec{\rho} &(\text{$\bm{x}$ is inside of the wall attached to $C$}).
\end{cases}
\end{equation}
This kink plays the role of the domain wall for the spontaneously-broken discrete chiral symmetry, and the energy densities inside and outside the kink are the same. 
Along the direction shown in Figure~\ref{fig:double-string-su_n}, the upper and lower kinks have the jump 
\begin{align}
    (\Delta \vec{\sigma})_{\mathrm{upper~kink}} &= \frac{2 \pi \vec{\rho}}{N}, \\
    (\Delta \vec{\sigma})_{\mathrm{lower~kink}} &= -\frac{2 \pi \vec{\rho}}{N} + 2 \pi \vec{\mu}_1, 
\end{align}
respectively. 
Let us confirm that this configuration satisfies the boundary condition. 
The winding number around the Wilson loop is correctly given by 
\begin{align}
    \oint_{C'}\diff \vec{\sigma}=(\Delta \vec{\sigma})_{\mathrm{upper~kink}}+(\Delta \vec{\sigma})_{\mathrm{lower~kink}}=2\pi \vec{\mu}_1. 
\end{align} 
Moreover, the twisted boundary condition requires\footnote{The negative sign on the right-hand-side arises from the difference of directions of the arrows in Figure~\ref{fig:double-string-su_n}.
Note that the difference $(\Delta \vec{\sigma})_{\mathrm{lower~kink}} = - \frac{2 \pi \vec{\rho}}{N} + 2 \pi \vec{\mu}_1$ is consistent with the configuration because of the weight-vector periodicity of the dual photon. Indeed, we have $(\vec{\sigma})_{\mathrm{inside}} = (\vec{\sigma})_{\mathrm{outside}} + (\Delta \vec{\sigma})_{\mathrm{lower~kink}} = - \frac{2 \pi \vec{\rho}}{N} + 2 \pi \vec{\mu}_1 \sim - \frac{2 \pi \vec{\rho}}{N}$, where we have used the weight-vector periodicity in the last line.} 
\begin{align}
    (\Delta \vec{\sigma})_{\mathrm{lower~kink}}|_{x_3=0}=P_W (-\Delta \vec{\sigma})_{\mathrm{upper~kink}}|_{x_3=L_3},
\end{align}
and we can see that this relation holds because of $P_W \vec{\rho} = \vec{\rho} - N \vec{\mu}_1$.

Summing up these two contributions, we obtain the expectation value of the Wilson loop as 
\begin{equation}
    \langle W(C)\rangle \sim \rme^{-T_{\mathrm{conf}}\, \mathrm{Area}(C)} + \rme^{- T_{\mathrm{kink}}\, L_3\, \mathrm{Length}(C)}, 
\end{equation}
and the second term dominates when $C$ becomes sufficiently large. 
This provides an explanation from the $3$d perspective on why the $2$d Wilson loop in the center-twisted $\mathbb{R}^2\times (S^1)_3$ obeys the perimeter law and generates the spontaneously-broken chiral symmetry. 
Therefore, the 2d perimeter law from the double-string picture also works well in the $SU(N)$ case.


\section{Bion amplitude from the quasi-moduli integration}
\label{sec:BionContribution}

In this section, we compute the interaction between monopoles and anti-monopoles to investigate how the bion amplitude transforms between $\RR^3 \times (S^{1})_4$ and  $\RR^2 \times (S^{1})_3 \times (S^{1})_4$ with 't Hooft flux when the size $L_3$ is gradually changed.

\subsection{Review on the bion amplitude of QCD(adj) in \texorpdfstring{$\RR^3 \times S^{1}$}{R3xS1}}

We first review the semiclassical physics of $n_f$-flavor $SU(N)$ QCD(adj.) on $\RR^3 \times (S^{1})_4$ with $NL_4\Lambda \ll 1$ following Refs.~\cite{Unsal:2007vu,Unsal:2007jx}. 
When $n_f\ge 2$, the one-loop GPY potential prefers the center-symmetric holonomy, $\vec{\phi}=\vec{\phi}_c$, and the 3d perturbative effective action for the diagonal components of gluons and fermions is
\begin{align}
S_{\rm eff} = \int_{\RR^3} \diff^3x \frac{L_4}{g^2} \Bigg[ \frac{1}{4}({\vec{f}_{ij}})^2 + 
\sum_{f=1}^{n_f}\vec{\Bar{\lambda}}_f  \sigma_i \partial_i  \vec{\lambda}_f \Bigg]\,,
\end{align} 
with $i,j=1,2,3$ denoting the three spatial directions. 
Associated with the adjoint higgsing, there appear $N$ types of fundamental monopoles ($N-1$ BPS and one KK). 
On $\RR^3$ each monopole carries $2n_f$ adjoint zero modes ($2$ per flavor), as dictated by the Callias index theorem. 
It is notable that the number of adjoint zero modes on each monopole equals $2n_f$ for the particular case of a $\ZZ_N$-symmetric Polyakov-loop background \cite{Misumi:2014raa}. 

The leading contribution to the bosonic potential comes from the bions, which are the bound state of monopoles and anti-monopoles so that fermionic zero modes are lifted. 
In Section~\ref{sec:MonopoleSemiclassics}, we obtained the bion amplitude using the power of supersymmetry as it can be related to the monopole amplitude. 
We can also compute the bion amplitude based on the 3d effective theory more directly by evaluating the interaction between monopoles and anti-monopoles~\cite{Unsal:2007jx,Anber:2011de,Argyres:2012ka}: The expression for the bion amplitude has the following structure, 
\begin{align}
Z_{\rm bion}(g) & \sim \underbrace{g^{-8}}_{\text{Jacobian}} \underbrace{\rme^{- 2(S_I/N) (1+cg)}}_{\text{Boltzmann weight}}
\int \diff^3x \diff^3y ~ \underbrace{\exp\left(-V_{\rm b} (\vec{x}-\vec{y})\right)}_{\text{dual photon exchange}} \underbrace{\big[ S_F(\vec{x}-\vec{y}) \big]^{2n_f}}_{\text{fermion exchange}} \,.
\label{eq:bioz}
\end{align}
The factor $g^{-8}$ appears from the Jacobian for collective coordinates, $\rme^{-2(S_I/N)(1+cg)}$ is the Boltzmann weight for two monopoles with $c = \sqrt{\frac{n_f -1}{3}}$ for $N=2$, and the rest comes from the interaction between the monopole and anti-monopole. 
$\exp(-V_{\rm b}(\vec{x}-\vec{y}))$ is the Coulomb interaction between monopoles located at $\vec{x}$ and $\vec{y}$ due to the dual-photon exchange, and its concrete form for $N=2$ is given by 
\begin{align}
V_{\rm b}(\vec{r})
= \pm\frac{4\pi L_4}{g^2} \frac{1}{r} \,,
\end{align} 
where we take $+$ for the magnetic bion (BPS-$\overline{\mathrm{KK}}$) and $-$ for the neutral bion (BPS-$\overline{\mathrm{BPS}}$). 
The last factor $\big[ S_F(\vec{x}-\vec{y}) \big]^{2n_f}$ comes from the fermion zero-mode exchange interaction with the massless free fermion propagator  
\begin{align}
S_F(\vec{r}) 
&=\lim_{m\to 0}  (-\im \sigma_i \partial_i) \int  \frac{\diff^3 k}{(2\pi)^3} \frac{\ee^{i \vec{k}\cdot \vec{r}}}{ \vec{k}^2 + m^2}
\nonumber\\ 
& = \frac{\im}{4\pi} \frac{\sigma_i x_i}{r^3} \,, 
\end{align}
where the higher Matsubara frequencies are omitted.
For $n_f =1$, i.e., for $\mathcal{N}=1$ SYM, the Coulomb potential has to be doubled, $V_b \rightarrow 2 V_b$, since the holonomy also gives the same long-range interaction.

From now on, we focus on $N=2$.
With an approximation $S_F(\vec{x})\sim 1/r^2$, the magnetic bion amplitude becomes 
\begin{align}
 Z_{\rm bion} & \sim \frac{1}{g^8}\exp\left(- \frac{8\pi^2}{g^2}(1+cg) \right) \int_{\rho_{\min}}^\infty \diff r  r^2 \exp\left[-V_{\rm eff}(r)\right]\,, 
  \label{eq:Zbion0}
\end{align}
where $\rho_{\min}\sim L_4$ is a cutoff of the $3$d effective description, and   
\begin{align}
V_{\rm eff}(r) = 
\frac{4\pi L_4 }{g^2 r} + 4 n_f \log \frac{r}{L_4} \, . 
\label{eq:veff0}
\end{align}
The minimum of $V_{\rm eff}(r)$ is located at $r=r_{\rm b}$ with $r_{\rm b} = \frac{\pi L_4 }{g^2 n_f}$. 
For $\mathcal{N}=1$ SYM, the bion size has an extra factor $2$ as $r_b=2\pi L_4/g^2$ because the Coulomb potential is doubled. We can identify it as the typical size of magnetic bions, and this implies that the magnetic bion is a shallow bound state as $r_{\rm b}\gg L_4$. 
By evaluating the integral \eqref{eq:Zbion0} with a Gaussian approximation, we find
\begin{align}
Z_{\rm bion}  \sim \frac{1}{g^{14-8n_f}}\exp\left(- \frac{8\pi^2}{g^2}(1+cg) \right)   \,.
\label{eq:Zbion01}
\end{align}
By substituting the two-loop coupling constant with
$\beta_0 = (22-4n_f)/3$ and $\beta_{1} = (136-64n_{f})/3$,
the mass gap is obtained as
\begin{align}
\frac{\cal M}{\Lambda} = 4\pi\sqrt{\frac{8 Z_{\rm bion}(g)}{g^2 L_4^2}} \sim (\Lambda L_4)^{\frac{8-2n_f}{3}}\,.
\end{align}
As this produces the nonperturbative mass gap for the dual photon, this result shows the microscopic dynamics behind the ``bion confinement mechanism" in QCD(adj) on $\RR^3 \times S^{1}$.

\subsection{Bion amplitude of \texorpdfstring{$\mathcal{N}=1$ SYM}{N=1 SYM} in \texorpdfstring{$\RR^2 \times S^{1} \times S^{1}$}{R2xS1xS1} with 't Hooft flux}

In what follows, we extend the previous computation of the bion amplitude on $\mathbb{R}^3\times (S^1)_4$ to the $\mathbb{R}^2\times (S^1)_3\times (S^1)_4$ setup with the 't Hooft flux and $NL_4\Lambda\ll 1$. For simplicity of calculations, let us focus on $\mathcal{N} = 1$ SYM, or $n_f = 1$ QCD(adj), with the gauge group $SU(2)$.
We use $(x_1, x_2)$ as the coordinate for $\RR^2$ and $x_3\sim x_3+L_3$ for $(S^{1})_3$. 
We only take into account the zero modes along the $x_4$ direction as $L_4$ is sufficiently small. 

Because of the twisted boundary condition, $\sigma(x_3+L_3)=-\sigma(x_3)$, the Coulomb potential between monopoles is
\begin{align}
V_{\rm b}(\vec{r})
&= \pm \frac{16\pi^2  }{g^2} 
\frac{L_4}{L_3}\sum_{n_3 \in 2\mathbb{Z}+1} \int  \frac{\diff k_1 \diff k_2}{(2\pi)^2} \frac{\ee^{\im \vec{k}\cdot \vec{r} }}{ k_1^2 +  k_2^2 + \omega^2 n_3^2}
\nonumber \\
 &= \pm\frac{16\pi}{g^2} \frac{L_4}{L_3}\sum_{n=1}^{\infty} K_{0}((2n-1)\omega\rho) \cos((2n-1)\omega x_{3}) \,, \label{eq:Coulomb_Fourier}
\end{align}
where we define $\omega \equiv \frac{2\pi}{NL_3} = \frac{\pi}{L_3}$, $\rho = \sqrt{x_1^2 + x_2^2}$ and $\vec{k} = (k_1,k_2,\omega n_3)$. $K_0$ is the modified Bessel function of the second kind. 
The fermion propagator for this case is
\begin{align}
S_F(\vec{r}) &=  \frac{1}{L_3}\sum_{n_3 \in 2\mathbb{Z}+1} \int  \frac{\diff k_1 \diff k_2}{(2\pi)^2} \frac{\ee^{i \vec{k}\cdot \vec{r}}}{\sigma_1 k_1 + \sigma_2 k_2 + \sigma_3 \omega n_3 }
\nonumber\\
& =\frac{1}{L_3 \pi}(-\im \sigma_i \partial_i) \sum_{n=1}^{\infty}  K_{0}((2n-1)\omega\rho) \cos((2n-1)\omega x_{3})\,.  \label{eq:Fermion_prop_Fourier}
\end{align}
We put the $\pm$ sign in front of the Coulomb potential $V_{\rm b}$ to distinguish the magnetic and neutral bions, but this turns out to be unnecessary for the $\mathbb{R}^2\times (S^1)_3\times (S^1)_4$ setup: 
It is important to note that the neutral bion and magnetic bion configurations are unified by extending the $x_{3}$ region from $[0,L_3)$ to $[0,N L_3) = [0, 2L_3)$. 
Remember that the $(\mathbb{Z}_N^{[0]})_{3\mathrm{d}}$-twisted boundary condition exchanges BPS monopole and KK monopole, and the species of monopole are no longer discriminated for the extended $x_3$ moduli. 
More concretely, one can confirm it by noticing that $V_{\rm b}(\rho, x_3+L_3)=-V_{\rm b}(\rho, x_3)$. 
In the following, we then take the $-$ sign for $V_{\rm b}$ without loss of generality.

The bion amplitude can be written as
\begin{align}
 Z_{\rm bion} & \sim \frac{1}{g^8}\exp\left(-  \frac{8\pi^2}{g^2}\right) \int_{\rho_{\min}}^\infty \diff\rho \int_{0}^{2L_3}\diff x_3~ \rho\exp\left(-V_{\rm eff}(\rho,x_3)\right)\,, 
  \label{eq:Zbion3}
\end{align}
with the effective potential
\begin{align}
    V_{\mathrm{eff}}(\rho, x_3) = 2 V_{\mathrm{b}}(\rho, x_3) - \log \left( (\partial_3 V_{\mathrm{b}})^2 + (\partial_\rho V_{\mathrm{b}})^2 \right).
    \label{eq:Veff_rx3}
\end{align}
The prefactor $2$ for $V_{\rm b}$ arises from the contribution of the holonomy degrees of freedom, and the second term describes the fermion-exchange contribution.  
Let us again emphasize that this bion amplitude (\ref{eq:Zbion3}) includes both magnetic bion and neutral bion, because of the extended moduli integral $x_3 \in [0, 2L_3)$. 
In our convention for choosing of the $-$ sign, the neutral bion exists near $x_3 \approx 0$ and the magnetic bion exists near $x_3 \approx L_3$, when $L_3$ is supposed to be sufficiently large.

For later purposes, let us derive the asymptotic form in the long-distance region\footnote{As we will see below, the effective potential in this limit becomes relevant when $L_4 \ll L_3 \ll L_4/g^2$.} $\rho \gg \omega^{-1}$.
In this limit, we just take the $n=1$ mode into account in the expressions of the Coulomb potential (\ref{eq:Coulomb_Fourier}) and fermion propagator (\ref{eq:Fermion_prop_Fourier}).
Then, the Coulomb potential and the fermion propagator are reduced to
\begin{align}
V_{\rm b}(\rho, x_3) &\approx -\frac{16\pi }{g^2 } \frac{L_4}{L_3} K_{0}(\omega\rho) \cos(\omega x_{3}) \notag \\
&\approx -\frac{16\pi }{g^2 } \frac{L_4}{L_3} \sqrt{\frac{\pi}{2 \omega\rho}} \rme^{- \omega\rho} \cos(\omega x_{3}) \,,
\end{align}
and 
\begin{align}
|S_F(\rho, x_3) |^2
&\sim  
(K_{1}(\omega\rho))^2 \cos^2 (\omega x_3) + (K_{0}(\omega\rho))^2 \sin^2(\omega x_{3})
\nonumber\\
&\sim \frac{1}{2\omega \rho}\exp(- 2\omega \rho),
\end{align}
where we have used the asymptotic form of the modified Bessel function $K_\nu (z) \rightarrow \sqrt{\frac{\pi}{2z}} \rme^{-z}$ as $z \rightarrow \infty$.
This exponential falloff reflects the mass gap $\omega =  \frac{\pi}{L_3}$ arising from the 't Hooft twist. 
In this limit, the bion amplitude reads
\begin{align}
 Z_{\rm bion} & \sim \frac{1}{g^8}\exp\left(-  \frac{8\pi^2}{g^2}\right) \int \diff \rho \int_{0}^{2L_3}\diff x_3~ \exp\left(\frac{32 \pi }{g^2 } \frac{L_4}{L_3} \sqrt{\frac{\pi}{2 \omega\rho}} \rme^{- \omega\rho} \cos(\omega x_{3}) - 2 \omega \rho \right)\,. \label{eq:large-rho-amplitude}
\end{align}
We note that this amplitude is similar to the bion quasi-moduli integral in the ${\mathbb Z}_{N}$-twisted ${\mathbb C}P^{N-1}$ sigma model on $\RR \times S^1$ \cite{Fujimori:2016ljw}.
Here, we can integrate out the $x_3$-moduli integral, and we have the intermediate expression of the integral 
$ \int_{\rho_{\min}}^\infty \diff \rho \int_{0}^{2L_3}\diff x_3 ...$ as
\begin{align}
 &\int_{\rho_{\min}}^\infty \diff \rho \,
I_0 \left[\frac{32 \pi }{g^2 } \frac{L_4}{L_3} \sqrt{\frac{\pi}{2 \omega\rho}} \rme^{- \omega\rho}\right]\, \ee^{-2\omega\rho}  \,,
\end{align}
with the modified Bessel function $I_0$. The integrand of this amplitude can be interpreted as the 2d effective potential between the center vortex and anti-vortex, 
\begin{align}
{\tilde V}_{\rm eff}(\rho) = 
-\log\left( I_0  \left[\frac{32 \pi }{g^2 } \frac{L_4}{L_3} \sqrt{\frac{\pi}{2 \omega\rho}} \rme^{- \omega\rho}\right] \right)  + 2 \omega \rho \,. 
\end{align}

\subsection{Change of magnetic-bion saddle}

In $\mathbb{R}^3 \times S^1$, the magnetic bion appears as a minimum of the effective potential of the quasi-moduli integral of the BPS-$\overline{\mathrm{KK}}$ molecule, (\ref{eq:Zbion0}).
Now, let us observe how the magnetic-bion saddle changes in the $\mathbb{R}^2 \times S^1 \times S^1$ setup. 

When we are interested in the scale $\rho \sim O(L_3)$, it would be efficient to use the original expressions for the Coulomb potential, which is derived from the mirror-image configuration (Figure \ref{fig:monopole-vortex-schematic}).
In this case at $N=2$, we have the Coulomb potential,
\begin{align}
    V_{\mathrm{b}}(\rho, x_3) = - \frac{4 \pi L_4}{g^2 L_3} \sum_{m \in \mathbb{Z}} \left\{ \frac{L_3}{\sqrt{(x_3 - 2 m L_3)^2 + \rho^2}} - \frac{L_3}{\sqrt{(x_3 - (2 m + 1) L_3)^2 + \rho^2}}  \right\}. 
    \label{eq:Coulomb_pot_for_figure}
\end{align}
In what follows, we consider the effective potential (\ref{eq:Veff_rx3}) with the above $V_{\mathrm{b}}(\rho, x_3)$.
With the extended moduli $x_3\sim x_3+2 L_3$, the neutral bion, BPS-$\overline{\mathrm{BPS}}$, is around $(\rho,x_3) \approx (0,0)$, and the magnetic bion, BPS-$\overline{\mathrm{KK}}$, is around $(\rho,x_3) \approx (0,L_3)$.

We now study the saddle-point structure of the effective potential numerically.
The important dimensionless parameter in the effective potential is $\frac{4 \pi L_4}{g^2 L_3}$, which is the ratio between the size of $(S^1)_3$ and the 3d bion scale $\sim L_4/g^2$. 
For the numerical purpose, we introduce the cutoff to the sum of the Coulomb potential (\ref{eq:Coulomb_pot_for_figure}):
\begin{align}
    \sum_{m \in \mathbb{Z}} \Bigl\{ \cdots \Bigr\} \Rightarrow \sum_{m = - N_{\mathrm{cutoff}}}^{N_{\mathrm{cutoff}}} \Bigl\{ \cdots \Bigr\} .
\end{align}
In the figures below, we use $N_{\mathrm{cutoff}} = 30$.

\begin{figure}[t]
\centering
\begin{minipage}{.47\textwidth}
\includegraphics[width= \textwidth]{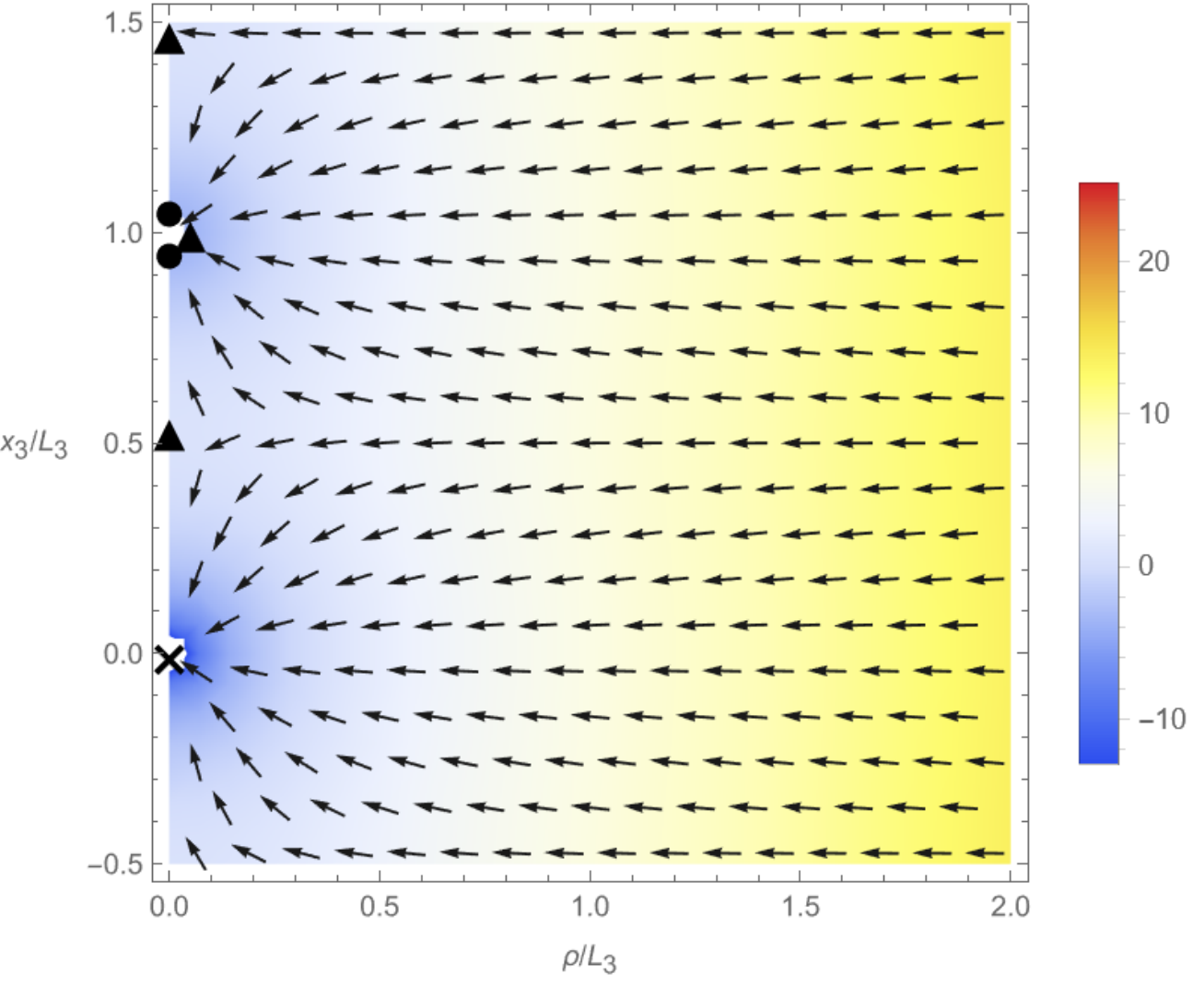}
\end{minipage}\quad
\begin{minipage}{.47 \textwidth}
\includegraphics[width= \textwidth]{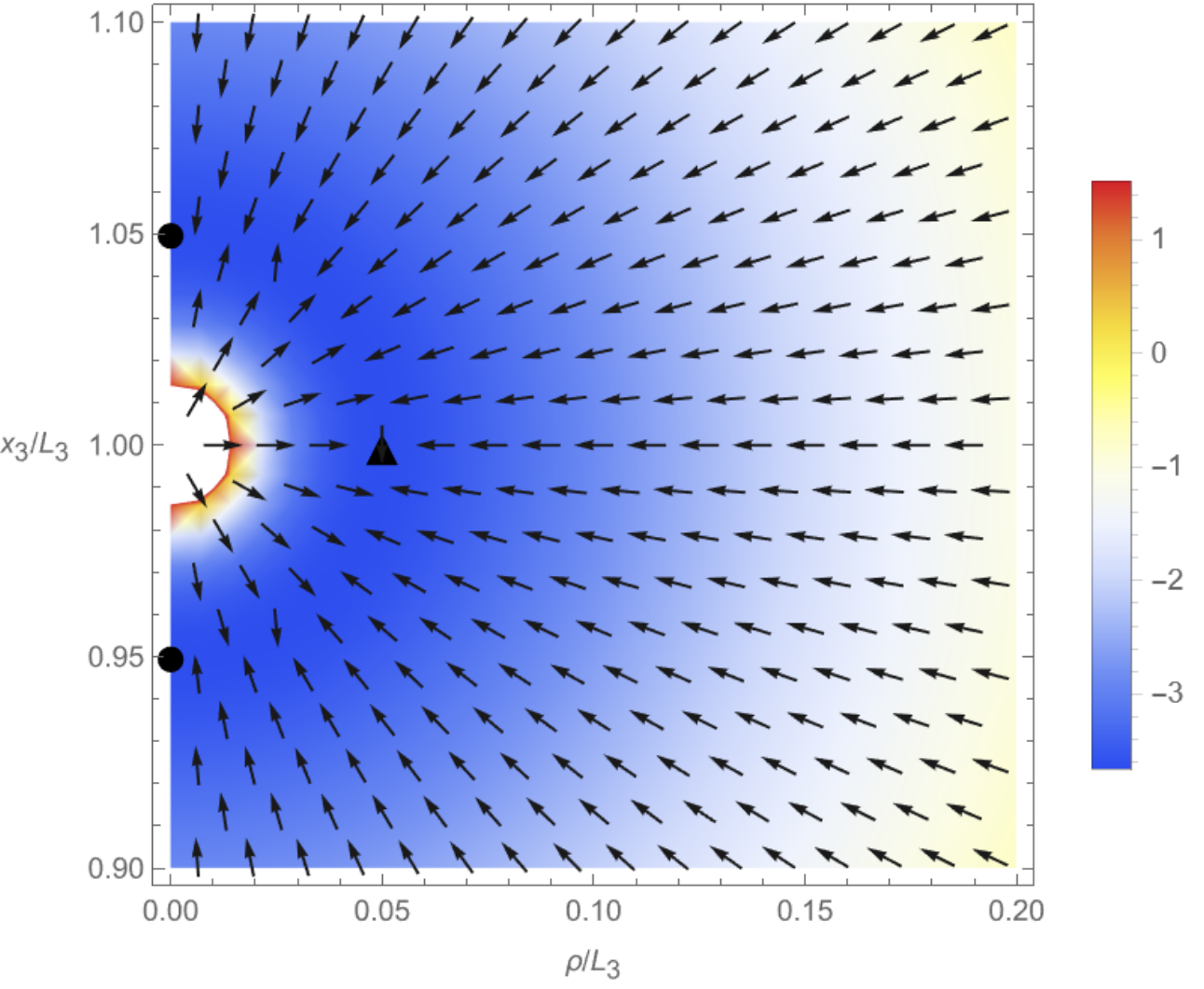}
\end{minipage}
\caption{The effective potential at $\frac{4\pi L_4}{g^2 L_3} = 0.1$: The heatmap describes the height of the effective potential, and the arrows indicate the downward flows. The circle and triangle markers show locations of the saddle points (the circle marker represents a local minimum, and the triangle one corresponds to an unstable saddle). 
(\textbf{Left panel}) The plot in the region ($0< \rho /L_3< 2$, $-0.5 < x_3/L_3< 1.5$). The cross marker at $(\rho,x_3) = (0,0)$ denotes the BPS-$\overline{\mathrm{BPS}}$ singularity.
There are several saddle points around the magnetic-bion point $(\rho,x_3) \approx (0, L_3)$.
The unstable saddles near $(\rho,x_3) \approx (0, 0.5 L_3),~(0, 1.5 L_3)$ are akin to the “saddles at infinity.”
(\textbf{Right panel}) The magnified figure near the magnetic-bion point.
From this figure, it can be observed that the potential valley lies at $\rho^2 + (x_3-L_3)^2 \approx r_b^2$, which reflects the 3d spherical symmetry for the magnetic bion saddles in $\mathbb{R}^3 \times S^1$.
}
\label{fig:bion-potential-01}
\end{figure}

As an illustration of the effective potential at large $L_3 \gg L_4/g^2$, the saddle-point structure of the effective potential at $\frac{4\pi L_4}{g^2 L_3} = 0.1$ is depicted in Figure \ref{fig:bion-potential-01}.
In this figure, the color represents the values of the effective potential, and the arrows denote the downward flows.
Saddle points are marked with a circle for local minima and a triangle for unstable saddles.
The left figure is a plot in the region ($0< \rho /L_3< 2$, $-0.5 < x_3/L_3< 1.5$), and the right figure offers a zoomed view of the area $0< \rho /L_3<0.2$, $0.9< x_3/L_3<1.1$, close to the magnetic-bion point $(\rho,x_3) = (0,L_3)$.
We can observe the following behaviors of the effective potential:
\begin{itemize}
    \item BPS-$\overline{\mathrm{BPS}}$ singularity near the origin $(\rho,x_3) = (0,0)$ and the neutral bion. 

    Both the magnetic Coulomb interaction and the fermion exchange interaction are attractive around $(\rho,x_3) = (0,0)$, and there is no saddle point in the real domain. Moreover, the limit $(\rho,x_3)\to 0$ is singular but we should note that $\sqrt{\rho^2+x_3^2}\lesssim L_4$ is outside the scope of the effective theory. 
    In the complexified domain, there are saddle points at $\sqrt{\rho^2+x_3^2}\approx \rme^{\pi \im} r_{\rm b}$ with $r_{\rm b}=2\pi L_4/g^2$, and they correctly reproduce the neutral-bion amplitude including its sign.  
    
    \item Magnetic bion (BPS-$\overline{\mathrm{KK}}$) saddles around $(\rho,x_3) =(0, L_3)$ 

    Around $(\rho,x_3) = (0, L_3)$, the Coulomb interaction becomes repulsive, while the fermion-exchange interaction is attractive. 
    As a result, the potential has a valley at $\rho^2+ (x_3-L_3)^2  \approx r_{\rm b}^2$, which is identified as the 3d magnetic bion formation.
When $L_3$ is infinite, all the points of the potential valley are stable saddles and exactly degenerate due to the $3$d rotational symmetry. 
The degeneracy is slightly lifted by the finite $L_3$ effect, and local minima are found on the $x_3$ axis around $(\rho, x_3)\approx (0, L_3\pm r_{\rm b})$, and unstable saddles appear around $(\rho,x_3)\approx (r_{\rm b}, L_3)$ (see the right panel of Figure~\ref{fig:bion-potential-01}).

    \item Unstable saddles at $(\rho,x_3) \approx (0,\pm 0.5 L_3) $

    Far from the Coulomb core, the fermion-exchange attractive interaction is dominant.
    Near $(\rho,x_3) \approx (0, \pm 0.5 L_3)$, there is a point where the attractive force from BPS monopole at $x_3 = 0$ and the attractive force from KK monopole at $x_3 = L_3$ are balanced.
    This saddle resembles the ``saddle at infinity\footnote{The term ``saddle at infinity'' refers to an unstable saddle point that appears upon compactification, typically represented by an instanton–anti-instanton pair located at antipodal points.
    This saddle point is often related to the thimble of the complex neutral bion \cite{Behtash:2018voa}.}.'' 
\end{itemize}
This summarizes the situation at large $L_3$. In particular, the magnetic bion exists as local minima of the effective potential.

\begin{figure}[t]
\centering
\begin{minipage}{.31\textwidth}
\subfloat[$\frac{4\pi L_4}{g^2 L_3} = 0.2$]{
\includegraphics[width= \textwidth]{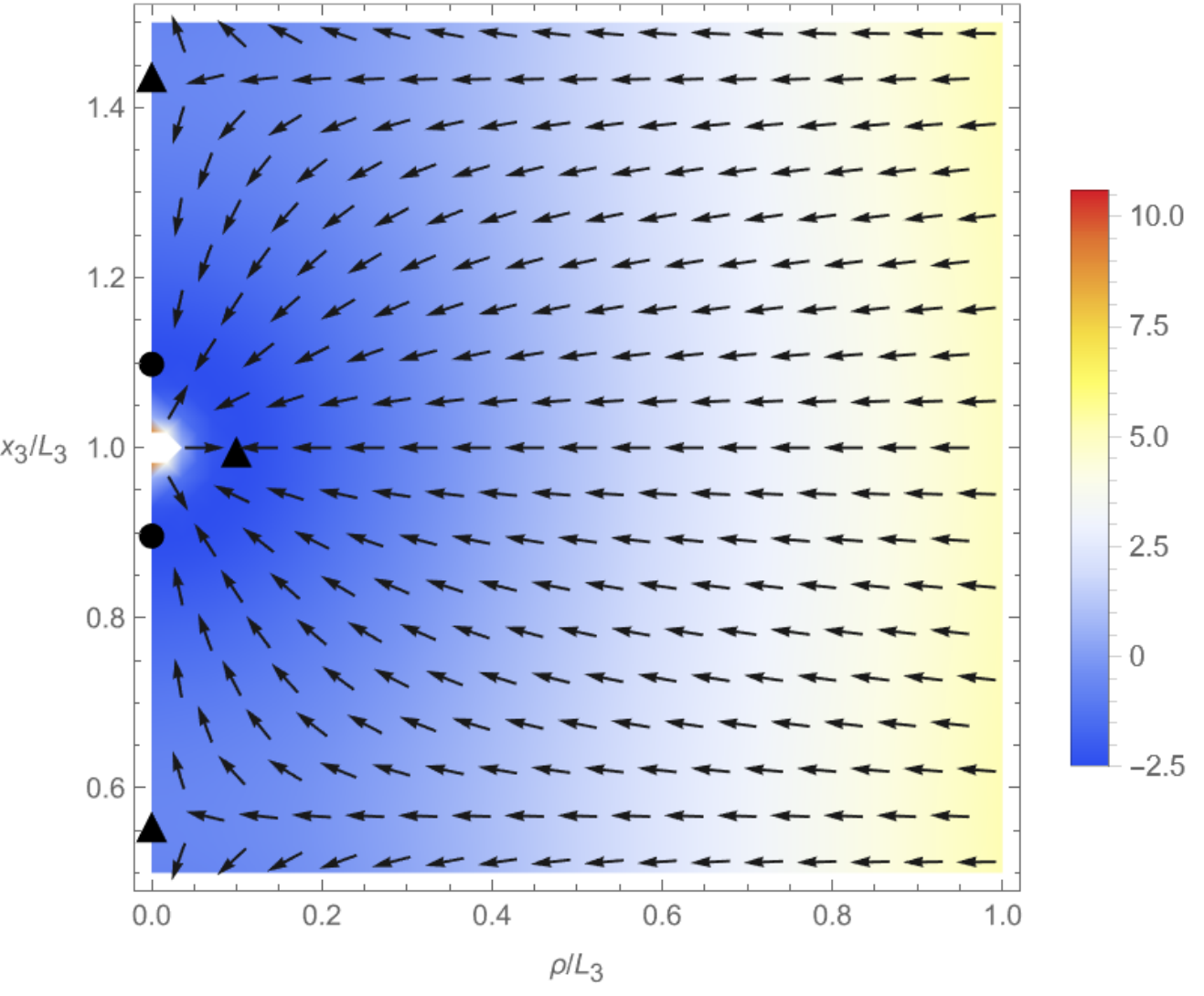}
\label{fig:bion_potential_02}
}\end{minipage}\quad
\begin{minipage}{.31\textwidth}
\subfloat[$\frac{4\pi L_4}{g^2 L_3} = 0.4$]{\includegraphics[width= \textwidth]{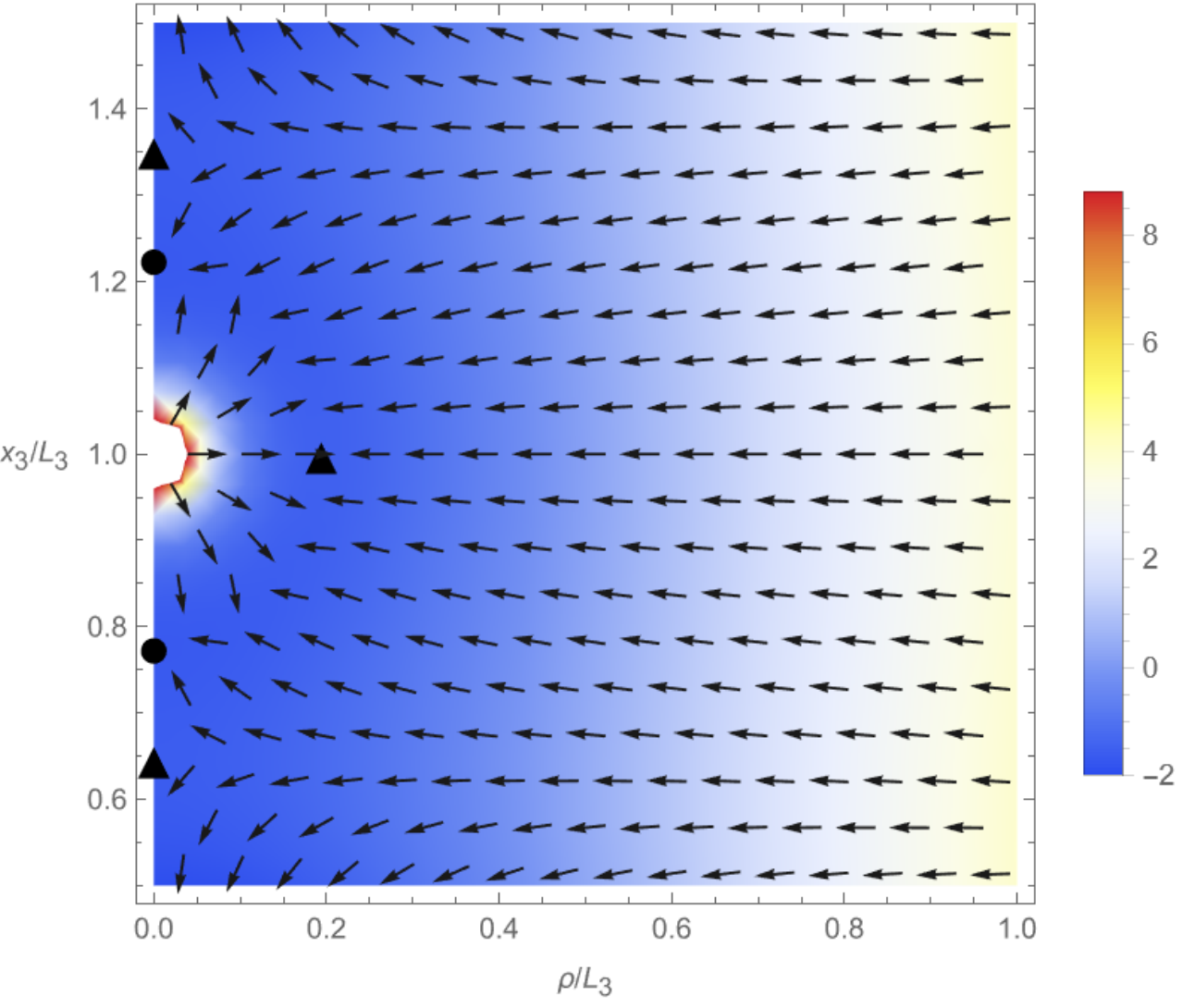}
\label{fig:bion_potential_04}}
\end{minipage}\quad
\begin{minipage}{.31\textwidth}
\subfloat[$\frac{4\pi L_4}{g^2 L_3} = 0.6$]{\includegraphics[width= \textwidth]{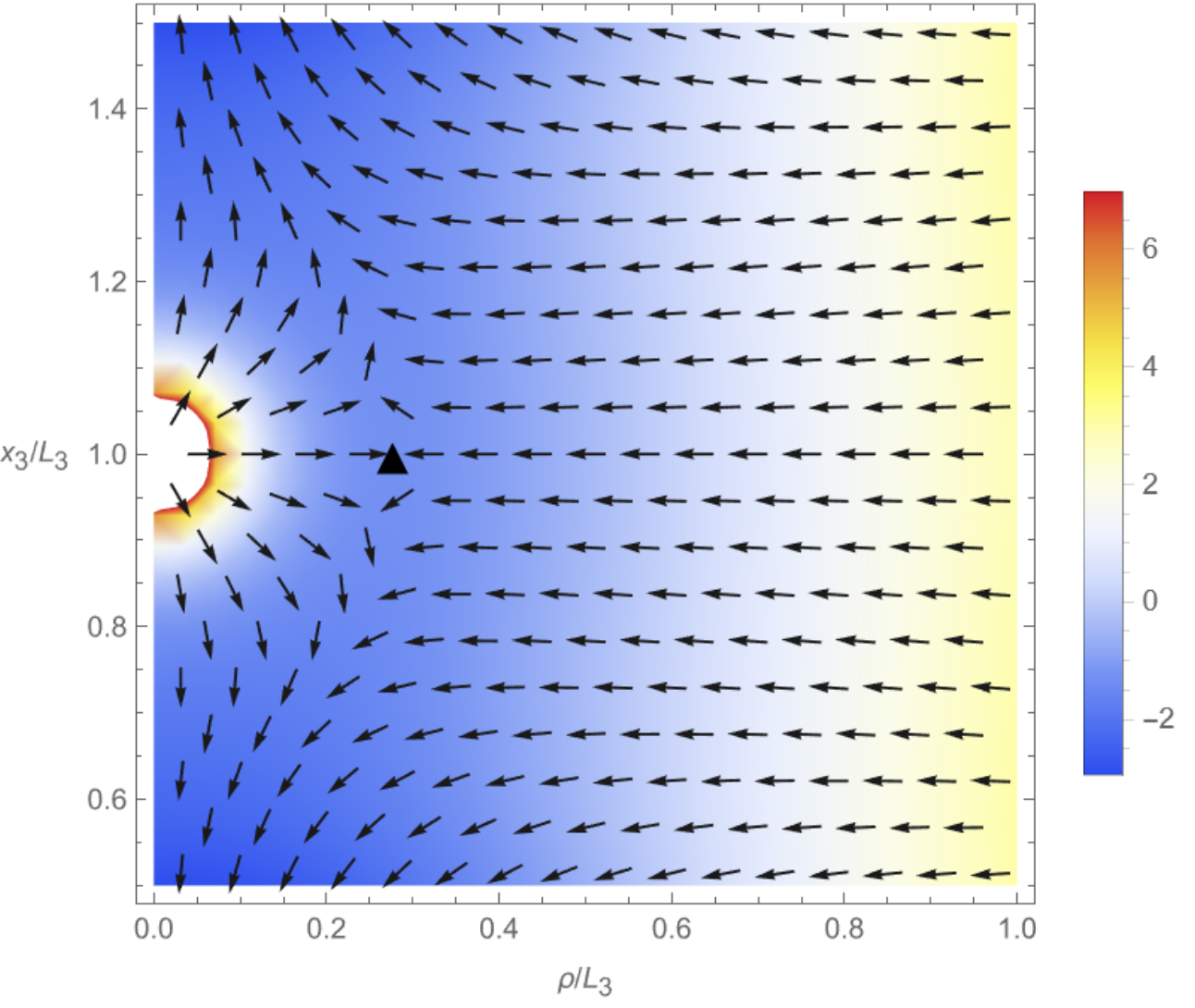}
\label{fig:bion_potential_06}}
\end{minipage}
\caption{Evolution of saddle-point structure of the effective potential in the magnetic-bion region $0.5 L_3 < x_3< 1.5 L_3$, as the parameter $\frac{4\pi L_4}{g^2 L_3}$ increases.
(\textbf{a}) The plot at $\frac{4\pi L_4}{g^2 L_3} =0.2$, where the situation is still similar to the large-$L_3$ case.
(\textbf{b}) The plot at $\frac{4\pi L_4}{g^2 L_3} = 0.4$. The local minimum and unstable saddle on the $x_3$ axis draw closer together as $\frac{4\pi L_4}{g^2 L_3}$ increases.
(\textbf{c}) The plot at  $\frac{4\pi L_4}{g^2 L_3} = 0.6$. As $\frac{4\pi L_4}{g^2 L_3}$ increases further, the local minimum and unstable saddle annihilate each other.
At this stage, only the unstable saddle at $x_3 = L_3$ remains.
}
\label{fig:saddle_point_structure}
\end{figure}

Then, let us see how the saddle-point structure changes when we make $L_3$ smaller, or $\frac{4\pi L_4}{g^2 L_3}$ becomes larger.
The structure in the neutral-bion region, $|x_3|< 0.5 L_3$, does not change qualitatively, so we will focus on the magnetic-bion region $|x_3 - L_3|< 0.5 L_3$.
This evolution is illustrated in Figure \ref{fig:saddle_point_structure}, which displays the saddle-point structures at $\frac{4\pi L_4}{g^2 L_3} = 0.2, 0.4, 0.6$.

An interesting phenomenon occurs on the $x_3$ axis.
At large $L_3$, e.g., Figure \ref{fig:bion_potential_02}, there are ``magnetic bion minima'' and ``(unstable) saddles at infinity'' on the $x_3$ axis.
As $L_3$ decreases, they move closer together (see the transition from Figure \ref{fig:bion_potential_02} to Figure \ref{fig:bion_potential_04}). 
Below the critical $L_3$, the local minima and unstable saddles merge, and they cancel each other out.
Eventually, only one unstable saddle remains (Figure \ref{fig:bion_potential_06}).
We can estimate the critical $L_3$ numerically, which is $\frac{4\pi L_4}{g^2 L_3} \approx 0.44$. 

This indicates that \textit{the magnetic bion no longer exists as local minima at small $L_3$}.
Then, following the path of this downward flow, one eventually reaches the singularity at the origin.
The only remnant of the magnetic bion is the unstable saddle at $(\rho, x_3) = (\rho_*,  L_3)$.
As the BPS/KK monopoles are unified as a single center vortex, the neutral and magnetic bions will become just a pair of center vortex and anti-vortex.
At small $L_3$, when we look at the microscopic structure of the $x_3$ internal moduli, the vortex–antivortex pair possesses a complex neutral bion saddle as well as the unstable saddle point, that is the remnant of the magnetic bion.
We interpret the disappearance of the stable magnetic-bion saddle as the microscopic signature of the crossover from the 3d monopole/bion picture to the 2d center-vortex picture.

\begin{figure}[t]
\centering
\begin{minipage}{.44\textwidth}
\includegraphics[width= \textwidth]{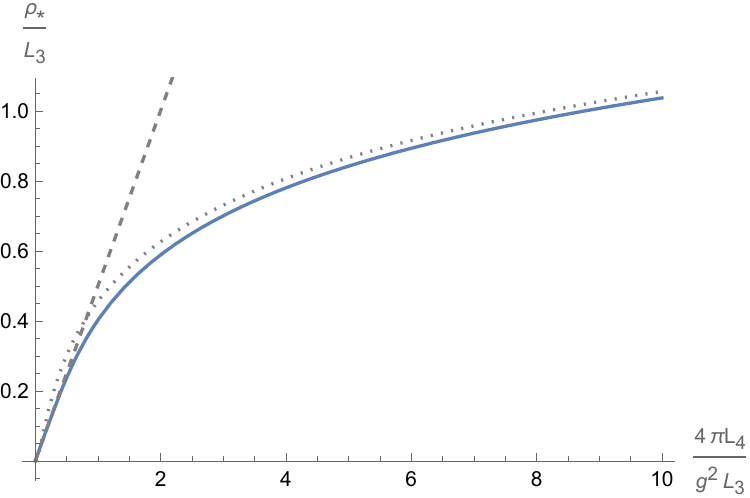}
\end{minipage}
\caption{ 
The $\frac{4\pi L_4}{g^2 L_3}$-dependence of the size of magnetic bion $\rho_*$, or the position of the unstable saddle $(\rho,x_3) = (\rho_*, L_3)$. We note that the vertical axis is taken as $\rho_*/L_3$ instead of $\rho_*$ itself. 
The solid curves show the numerical results, and the dashed and dotted curves are the asymptotic formula \eqref{eq:bion_location_asymptotic} for large-$L_3$ and small-$L_3$, respectively. }
\label{fig:bion-size}
\end{figure}

Before concluding this subsection, let us observe $L_3$ dependence of ``the size of magnetic bion'' $\rho_*$, or the location of the unstable saddle, which is plotted in Figure \ref{fig:bion-size}.
The asymptotic behaviors at small $L_3$ and large $L_3$ can be estimated as follows: 
\begin{align}
    \rho_* \simeq 
    \left\{
    \begin{array}{cc}
        \frac{2\pi L_4}{g^2} & \qquad (L_3\gg r_{\rm b}) , \\
        \frac{L_3}{2\pi} \, W_0 \left( 16 \pi \left( \frac{4 \pi L_4}{g^2 L_3} \right)^2 \right) & \qquad (L_3\ll r_{\rm b}), 
    \end{array}
    \right.
    \label{eq:bion_location_asymptotic}
\end{align}
where $W_0(x)$ is the principle branch of the Lambert $W$ function. For large $L_3$, the bion size should be determined by the one for the $3$d semiclassics, and for small $L_3$, the asymptotic form can be estimated from the large-$\rho$ expression (\ref{eq:large-rho-amplitude}).
These asymptotic forms are compared with the numerical result for $\rho_*/L_3$ in Figure \ref{fig:bion-size}. 

\subsection{On the hidden topological angle}

In this subsection, we study the interplay between the complex neutral bion and the unstable saddle (magnetic bion remnant).
We know that the vacuum energy vanishes in $\mathcal{N}=1$ SYM for supersymmetric vacua.
In the $\mathbb{R}^3 \times S^1$ monopole semiclassics, the magnetic bion decreases the vacuum energy but the neutral bion gives the opposite contribution, and they exactly cancel with each other and the supersymmetry is preserved. 
This relative sign for the neutral bion is called the hidden topological angle, which has been understood from the Bogomol'nyi--Zinn-Justin (BZJ) prescription for the quasi-moduli integral~\cite{Argyres:2012ka} and also from the complex saddles in its Lefschetz-thimble integration~\cite{Behtash:2015kna} (see also \cite{Nguyen:2023rww, Behtash:2015loa}). 

A natural question is how this cancellation is preserved during the reduction from 3d semiclassics to 2d semiclassics.
Here, we offer a few observations on this cancellation in the $\mathbb{R}^2 \times S^1 \times S^1$ setup.
First, let us see that some cancellation must happen in the quasi-moduli integral.
Due to supersymmetry \cite{Balitsky:1985in}, the integrand for the quasi-moduli integration of the bion amplitude satisfies the special property, 
\begin{align}
    \left( (\partial_3 V_{\mathrm{b}})^2 + (\partial_\rho V_{\mathrm{b}})^2 \right) \rme^{-2V_{\mathrm{b}}} = \frac{1}{4}\nabla^2  \rme^{-2 V_{\mathrm{b}}},
\end{align}
which follows from $\nabla^2 V_{\mathrm{b}} = 0$ up to singularity at the origin.
Therefore, the bion amplitude can be written as the integration of the total derivative:
\begin{align}
 Z_{\rm bion} & \sim \frac{1}{g^8}\exp\left(-  \frac{8\pi^2}{g^2}\right) \int_{\rho_{\min}}^\infty \diff \rho \int_{0}^{2L_3}\diff x_3~ \rho \left( (\partial_3 V_{\mathrm{b}})^2 + (\partial_\rho V_{\mathrm{b}})^2 \right) \rme^{-2 V_{\mathrm{b}}} \notag \\
 & = \frac{1}{4g^8}\exp\left(-  \frac{8\pi^2}{g^2}\right)  \int_0^{2L_3} \diff x_3 \int \diff \rho~ \rho \left( \partial_3^2 + \frac{1}{\rho} \partial_\rho \rho \partial_\rho \right)  \rme^{-2V_{\mathrm{b}}}.
  \label{eq:Zbion4}
\end{align}
After properly regularizing the neutral-bion singularity at $\rho=0$ (as in $\mathbb{R}^3 \times S^1$ case), this amplitude should vanish because of the exponential decay\footnote{In the $\mathbb{R}^3 \times S^1$ setup, the boundary terms of the neutral and magnetic bion at $|\vec{r}| \rightarrow \infty$ are finite and have opposite signs, thus canceling each other out.}: $V_{\mathrm{b}}(\rho, x_3) \sim \frac{1}{\sqrt{\rho}} \rme^{-\omega \rho}$.
At least, the total bion amplitude $Z_{bion}$ can be rewritten by only its boundary terms.
Thus, if there are relevant thimbles, they should be eventually canceled.

Next, let us try to understand the hidden topological angle in a more concrete way.
In the particular limit $L_4 \ll L_3 \ll L_4/g^2$, the location of the unstable bion saddle is given by the small-$L_3$ asymptotic formula of (\ref{eq:bion_location_asymptotic}).
Upon taking this limit, the size of magnetic bion becomes much larger than $L_3$: $\rho_* \sim \frac{L_3}{2\pi} \log \{16 \pi (4\pi L_4/g^2 L_3)^2\} \gg L_3$.
The quasi-moduli integral can be then written as (\ref{eq:large-rho-amplitude}), 
\begin{align}
 Z_{\rm bion} & \sim \int \diff \rho \int_{0}^{2L_3}\diff x_3~ \exp\left(\frac{32 \pi }{g^2 } \frac{L_4}{L_3} \sqrt{\frac{\pi}{2 \omega\rho}} \rme^{- \omega\rho} \cos(\omega x_{3} ) - 2 \omega \rho \right)\,. 
\end{align}
In this expression, the saddles can be easily found: 
\begin{itemize}
    \item The magnetic-bion unstable saddle:
\begin{align}
    (\rho, x_3) \simeq (\rho_*, L_3 ).
\end{align}
    \item The complex neutral-bion saddle:
    its location is approximately
    \begin{align}
    (\rho, x_3) \simeq (\rho_* \pm \im \pi, 0),
\end{align}
where we have used $\sqrt{\omega (\rho_* \pm \im \pi)} \simeq \sqrt{\omega \rho_*}$ from $\omega \rho_* \gg 1$.
\end{itemize}

We rewrite $\rho = \rho_* + \delta \rho$, and let us focus on the integration near the bion saddle(s) $|\delta \rho| \ll \rho_*$.
This approximation suffices for our purpose to observe the cancellation structure of these two saddles.
Then, the integration near the bion saddle can be expressed as\footnote{The coefficient is simplified by $\frac{16 \pi L_4}{ g^2  L_3}\sqrt{\frac{\pi}{2}} \frac{\exp\left( -\omega \rho_* \right)  }{\sqrt{\omega \rho_*}} = 1.$}
\begin{align}
     Z_{\rm bion} \propto \int_0^{2L_3} &\diff x_3 \int \diff (\delta \rho) ~ 
     \rme^{2 \rme^{- \omega \delta \rho }  \cos(x_3) - 2 \omega \delta \rho }. 
\end{align}
Interestingly, this integration is identical to that of the quasi-moduli integral of the supersymmetric $\mathbb{C}P^1$ quantum mechanics \cite{Misumi:2014jua,Misumi:2015dua}, whose thimble structure has been studied in \cite{Fujimori:2016ljw, Fujimori:2017oab}.
Depending on the infinitesimal phase to $g^2$, the intersection numbers of thimbles change.
Borrowing the result of Ref.~\cite{Fujimori:2016ljw}, we have
\begin{align}
     Z_{\rm bion} =
    \begin{cases}
        I_{\mathrm{mag}} - I_{\mathrm{neu, \rho_*-\im \pi }} ~~ (g^2 \rightarrow g^2 \rme^{- \im 0})\\
        I_{\mathrm{neu, \rho_* + \im \pi }} - I_{\mathrm{mag}}  ~~ (g^2 \rightarrow g^2 \rme^{+ \im 0})
    \end{cases}
    = 0,
\end{align}
where $I_{\mathrm{mag}}$ is the integration on the Lefschetz thimble from the magnetic bion saddle, and $I_{\mathrm{neu, \rho_* \pm \im \pi }}$ is that of the neutral bion saddle $(\rho, x_3) \simeq ( \rho_* \pm \im \pi,0)$.

These observations explain how the cancellation structure is preserved when we move from the 3d to the 2d semiclassical regime by changing $L_3$ with $L_4\ll L_3\ll L_4/g^2$.
The cancellation structure between real and complex saddles, or the hidden topological angle, remains intact nontrivially in the 2d vortex-antivortex framework.

\section{Summary and Discussion}

We studied ${\mathcal N}=1$ $SU(N)$ SYM theory using the weak-coupling semiclassical theory, and we made the connection between the monopole/bion-based semiclassics on $ {\mathbb R}^3 \times S^1 $ and the vortex-based semiclassics on ${\mathbb R}^2 \times T^2 $ with the 't Hooft twist. 
We showed that the BPS/KK monopoles in the $3$d effective theory are transmuted into the center vortex in the $2$d description across the dimensional reduction, and we have confirmed the ``weak-weak" continuity of the vacuum structure and the gluino condensate.

We also studied the behavior of the Wilson loop during the dimensional reduction from $ {\mathbb R}^3 \times S^1 $ to $ {\mathbb R}^2 \times T^2 $, and showed that the perimeter law of the 2d center-vortex semiclassics is understood from the 3d monopoles through the ``double-string picture''. Thanks to the center-twisted boundary condition, we uncovered that the domain wall wrapping along the compactified $S^1$ direction comes out of the Wilson loop, and the vacua inside and outside of the wall are related by the spontaneously broken chiral transformation. 
As a result, the Wilson loop in the $2$d semiclassics can be identified with the chiral-symmetry generator when the loop is sufficiently large, which is required to satisfy the anomaly matching condition with gapped vacua. 
Based on this observation, we also showed that the mass deformation, which breaks the discrete chiral symmetry, restores the area-law behavior of the Wilson loop.

\hl{On the one hand, we established the adiabatic continuity between the monopole and center-vortex semiclassical theories from $3$d to $2$d, with its explicit demonstration for dynamics, local observables, and vacuum structures. On the other hand, we also found the switching of the Wilson loop behaviors from the $3$d area law to $2$d perimeter law for finite $L_3$ as mentioned above. At the first sight, these two observations may seem to have a conflict. However, they are actually consistent because the switching from the $3$d area law to the $2$d perimeter law occurs only for asymptotically large Wilson loops, which are larger than the compactification size $L_3$. This is justified by estimating the switching size as $L_*\approx \frac{T_{\mathrm{conf}}}{T_{\mathrm{kink}}}L_3$. }

We also performed the computation of the bion amplitude on $\mathbb{R}^2\times S^1\times S^1$, and we found that the local minima corresponding to the magnetic bions disappear when $L_3$ becomes comparable to the magnetic bion size, $r_{\rm b}=2\pi L_4 /g^2$. 
Since the BPS/KK monopole becomes the identical center vortex under the twisted boundary condition, the magnetic bion becomes the vortex--anti-vortex pair, and it can no longer exist as the stable molecule for $L_3\lesssim 4.5 r_{\rm b}$. 
We also studied the role of neutral bions across the dimensional reduction and confirmed the vanishing of the vacuum energy in terms of the bion amplitude.

We suspect that SYM with tiny supersymmetry-breaking deformation on $\mathbb{R}^2\times S^1\times S^1$ with 't~Hooft twist provides a good testing ground for studying the resurgence structure of $4$d gauge theories. The resurgence theory relates the asymptotic nature of perturbative expansion with the nonperturbative contribution~\cite{Dunne:2013ada, Basar:2013eka, Cherman:2014ofa}. 
For this purpose, we first need to have the well-defined perturbative expansion but the presence of massless gluons causes various difficulties when we try to start this program on $\mathbb{R}^4$~\cite{tHooft:1977xjm}. 
On $\mathbb{R}^2\times S^1\times S^1$ with 't~Hooft flux, all the perturbative modes have nonzero KK mass, and then we can evade the difficulty associated with the infrared divergence, which makes the perturbative expansion well-defined~\cite{Yamazaki:2017ulc}. 
As another important ingredient to study the resurgence structure, we need to have a deep understanding of the nonperturbative objects, such as fractional instantons and their composites. 
While the analytic properties of the fractional instantons on $\mathbb{R}^2\times T^2$ with small $T^2$ are not much known, the BPS and KK monopoles on $\mathbb{R}^3\times S^1$ are quite well understood. 
Starting from the $3$d monopole/bion theory and taking the center-twisted compactification, we are able to study the bion amplitude as we have demonstrated for the SYM case in this paper (see \cite{Fujimori:2018kqp,Nishimura:2021lno} for the 2d cases). 

For SYM case, the perturbative series vanishes trivially due to supersymmetry, and we cannot proceed the resurgence program in the naive way. 
However, the asymptotic divergence should immediately appear with supersymmetry-breaking deformation, and this can be formally achieved by treating the fermion flavor $n_f$ as a continuous parameter. Then, the bion amplitudes with $n_f\not=1$ should start to have the ambiguity to cancel the Borel ambiguity of the asymptotic series. 
This structure, called Cheshire cat resurgence~\cite{Kozcaz:2016wvy, Dunne:2016jsr, Fujimori:2017osz,Dorigoni:2017smz, Dorigoni:2019kux}, would be a good starting point for the investigation of the resurgence structure in the $4$d gauge theories as some computations can be performed at the supersymmetric point.

\acknowledgments

The authors thank Mithat \"{U}nsal for the collaboration at the early stage of this work and for reading the manuscript.
The authors also appreciate the YITP-RIKEN iTHEMS conference ``Generalized symmetries in QFT 2024'' (YITP-W-24-15) and the YITP long-term workshop ``Hadrons and Hadron Interactions in QCD 2024'' (YITP-T-24-02) for providing the opportunities of useful discussions.  
This work was partially supported by Japan Society for the Promotion of Science (JSPS) Research Fellowship for Young Scientists Grant
No. 23KJ1161 (Y.H.), by JSPS KAKENHI Grant No. 23K22489 (Y.T.), 23K03425 (T.M.), 22H05118 (T.M.), and by Center for Gravitational Physics and Quantum Information (CGPQI) at Yukawa Institute for Theoretical
Physics.

\appendix 

\section{Conventions of roots and weights}
\label{appendix:RootWeightConvention}

Our convention for the roots and weights for $SU(N)$ is as follows.
Let $\{\vec{e}_n\}_{n=1}^{N}$ represent the orthonormal basis of $\mathbb{R}^N$.
\begin{itemize}
    \item The simple roots are given by $\vec{\alpha}_n=\vec{e}_n-\vec{e}_{n+1}$ $(n=1, \ldots, N-1)$.
    They satisfy $\vec{\alpha}_n \cdot \vec{\alpha}_m = 2 \delta_{nm} - \delta_{n,m\pm1}$. 
    The affine root vector is defined by $\vec{\alpha}_N := \vec{e}_N-\vec{e}_{1} = - (\vec{\alpha}_1 + \cdots +\vec{\alpha}_{N-1})$
    \item The fundamental weight can be expressed as $\vec{\mu}_n=\vec{e_1}+\cdots+\vec{e}_n-\frac{n}{N}\sum_{k=1}^{N}\vec{e}_k$ $(n=1, \ldots, N-1)$.
    This satisfies $\vec{\mu}_n\cdot \vec{\alpha}_m=\delta_{nm}$. 
    \item The weight vector of the defining representation can be written as:
\begin{equation}
    \vec{\nu}_1=\vec{\mu}_1,\,\, \vec{\nu}_2=\vec{\mu}_1-\vec{\alpha}_1,\,\,\ldots, \,\, \vec{\nu}_N=\vec{\mu}_1-\vec{\alpha}_1-\cdots - \vec{\alpha}_{N-1}\,.
\end{equation}
The following formulas are useful: $\vec{\alpha}_n=\vec{\nu}_n-\vec{\nu}_{n+1}$, $\vec{\nu}_n\cdot \vec{\nu}_m=- \frac{1}{N} + \delta_{nm}$.
    \item Using the basis $\{\vec{e}_n\}_{n=1}^{N}$, we define the cyclic Weyl permutation $P_W$ as $P_W \vec{e}_n = \vec{e}_{n+1}$; in particular $P_W \vec{\alpha}_n = \vec{\alpha}_{n+1~(\operatorname{mod}N)}$ and $P_W\vec{\nu}_n=\vec{\nu}_{n+1~(\operatorname{mod}N)}$.
\end{itemize}


\bibliographystyle{utphys}
\bibliography{./QFT,./refs}

\end{document}